\documentclass[a4paper,12pt]{article}
\usepackage{graphicx} 
\usepackage{jheppub}
\usepackage{float}
\usepackage{amsmath}

\newcommand{\be}{\begin{equation}}
\newcommand{\ee}{\end{equation}}
\usepackage{xcolor}
\definecolor{mdtRed}{rgb}{0.8,0.2,0.2}
\usepackage{booktabs}
\usepackage{tikz}
\usepackage{array}
\usepackage{makecell}
\usepackage{tcolorbox}
\usepackage{hyperref}
\hypersetup{colorlinks=true,
linkcolor=blue,
urlcolor=blue
}

\title{On flat space limits of $AdS_4$ black holes}
\author{Pedro Vicente Marto, Shradha Ramakrishnan and Stefan Vandoren}
\affiliation{Institute for Theoretical Physics, Utrecht University, 3584 CE Utrecht, The Netherlands}
\emailAdd{p.vicentemarto@uu.nl}
\emailAdd{s.ramakrishnan@uu.nl}
\emailAdd{s.j.g.vandoren@uu.nl}

\abstract{With a view towards understanding flat space holography, we study the flat space limit of asymptotically $AdS_4$ black holes in gauged $\mathcal{N}=2$ supergravity. Using holographic renormalisation, we carefully analyse the flat space limit of the on-shell action of various black holes in this setting and make explicit the states that contribute to the partition function in this limit. We find several features: large black holes lead to a divergence in the partition function which can be understood as a thermodynamic volume divergence. Meanwhile, semiclassically, small black holes reproduce the partition function of their flat space counterparts, and we clarify the mapping between BPS, (near-)extremal or non-extremal black holes on both sides for charged and rotating solutions. Finally, by studying a suitable flat limit of the BPS Kerr-Newman-$AdS_4$ black hole in M-theory, we obtain a charged BPS solution in flat space. On the dual side, we determine a Carrollian-like limit on the  known ABJM partition function on $S^1\times S^2$ at large $N$, and show that it reproduces the correct entropy of the BPS charged black hole in asymptotically flat space. }

\begin{document}

\maketitle

\section{Introduction}

What happens to black holes in AdS in the limit of vanishing cosmological constant? And if flat space holography can be described as a flat space limit of AdS/CFT, how are states in the latter description of quantum gravity mapped to the ones in its flat counterpart? In this paper, we start addressing these questions by studying the (semiclassical) bulk partition function of black holes in $AdS_4$, and their corresponding flat space limits. 

\subsection*{Motivation: Flat space holography from AdS/CFT}

Through the AdS/CFT correspondence, an asymptotically AdS black hole geometry should describe a state (or an ensemble thereof) in the dual CFT defined on the timelike boundary. On the other hand, the flat space limit of asymptotically AdS geometries is believed to be dual to the Carrollian limit of relativistic field theories, see e.g. \cite{Bagchi:2025vri,Ruzziconi:2026bix} for recent reviews. We should therefore attempt to define states in a Carrollian field theory dual to an asymptotically flat black hole geometry by placing it in AdS. The flat space limits of these geometries should therefore be described by a state in the Carrollian limit of the dual (CFT), now living on the null boundary of the bulk.

Some of these questions were addressed recently in \cite{Poulias:2025eck} in three spacetime dimensions,\footnote{References \cite{Hao:2025btl,Hao:2026cqm} explicitly map the generators of the symmetry algebra and the resulting states from (A)dS to flat space.} where in the flat limit the outer horizon of the BTZ black hole is blown up to null infinity where the dual Carroll field theory lives. Meanwhile, the inner horizon goes over to a cosmological horizon, whose entropy was matched with that of the Carroll limit of a large class of 2D CFT's \cite{Barnich:2012xq,Bagchi:2012xr}. In $AdS_4$ a distinction between small and large black holes appears. The scaling of the large horizon radius with the AdS radius, $\ell_{AdS}$, would lead us to predict a similar blow up of large black holes to null infinity, while the small black holes contribute to the ensemble in the flat limit. Other features such as (non-)extremality and BPS-ness of charged and rotating black holes are characteristic to four dimensions, and we will quantify how all of these affect the mapping of states between asymptotically AdS and flat space.


 Our strategy is as follows: we consider black holes in minimal and gauged $\mathcal{N}=2$ supergravity\footnote{Minimal supergravity in our paper means pure supergravity with a (negative) cosmological constant, while the term gauged supergravity we use here for  matter couplings and a scalar potential with an AdS vacuum.} and compute their on-shell actions via holographic renormalisation. The semiclassical partition function takes the form
\begin{equation}
    Z_{\text{AdS}_4}\sim e^{-I_{\text{$AdS_4$,on-shell}}}.
\end{equation}
We then take the bulk flat space limit by carefully sending $\ell_{AdS}\rightarrow\infty$ wherever the limit is well-defined. Then the flat space bulk partition function should take the form
\begin{equation}
    Z_{\text{flat}}\sim e^{-I_{\text{flat, on-shell}}}\,.
\end{equation}

By doing this systematically across several classes of black hole solutions, we clarify what is the  mapping between different classes of bulk contributions. This analysis is summarized in Section \ref{sec:summary}. Particular features we encounter are that certain BPS geometries in flat space can be seen as arising from the flat limit of extremal non-BPS geometries in $AdS_4$, both at the level of the metric and at the level of the semiclassical partition function. Meanwhile, BPS geometries in $AdS_4$ generically become singular in the flat limit, especially when they correspond to excitations above the magnetic AdS vacuum of $\mathcal{N}=2$ supergravity. A special case we study in detail is that of the BPS limit of the Kerr-Newman-$AdS_4$ black hole, which admits a particular flat space limit that leads to the BPS charged black hole in flat space. We note that whenever a BPS state in flat space is reproduced by the flat limit of a state in AdS, we observe an enhancement of preserved supercharges.

\subsection*{Dual limits and flat space holography}

\begin{figure}[h]\label{Scheme}
\begin{center}
\begin{tikzpicture}[
    box/.style={
        draw,
        rounded corners=8pt,
        minimum width=3.2cm,
        minimum height=1.5cm,
        align=center,
        font=\small
    },
    >=stealth,
    every node/.style={font=\small}
]

\node[box] (bulk) at (0,3.0) {
    Bulk partition\\
    function in $AdS_4$
};

\node[box] (flat) at (7,3.0) {
    4d flat bulk partition\\
    function
};

\node[box] (cft) at (0,0) {
    3d relativistic field\\
    theory partition function
};

\node[box] (carroll) at (7,0) {
    Carrollian partition\\
    function for 3d theory
};

\draw[->] (bulk.east) -- 
    node[above] {$\ell_{AdS}\to\infty$} 
    (flat.west);

\draw[->] (cft.east) -- 
    node[above] {$c\to0$} 
    (carroll.west);

\draw[<->] (cft.north) -- 
    node[left] {AdS/CFT} 
    (bulk.south);

\draw[<->] (carroll.north) -- 
    node[right] {Carrollian holography} 
    (flat.south);

\end{tikzpicture}
\end{center}
\caption{Schematic approach to Carrollian holography for partition functions by embedding it into AdS/CFT. In this diagram, and in most of this paper, the $S^7$ sphere dependence is suppressed but ultimately needs to be taken into account when taking the flat space limit.}
\end{figure}
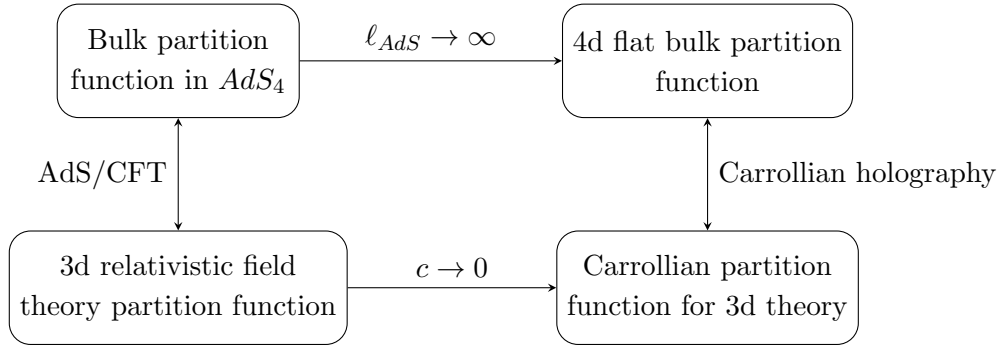

We also aim to take a step further and define suitable limits of field theory partition functions which would be dual to the flat space limits we study. This general programme is summarized in Figure \ref{Scheme}, where the suitable field theory limit would complete the bottom corner of this diagram. Doing this would require an exact knowledge of the partition function, at least to leading order in $N$ for which AdS/CFT is valid in the semiclassical gravity regime. We will therefore restrict our attempt at completing this diagram to a particular supersymmetric case. 

It is well known that BPS $AdS_4$ black holes can be described by the dual  $3d$ ABJM theory. This has been extensively studied in  (but not limited to) \cite{Marino:2011nm,Hristov:2013spa,Azzurli:2017kxo,Benini:2015eyy}. See also, for example, \cite{Marino:2011nm} for a computation of the partition function of ABJM on $S^3$. A concrete starting point in establishing the limit represented by the bottom arrow in Figure \ref{Scheme} would be to naively take the Carroll limit of ABJM. There are, however, subtleties in taking the Carroll limit of ABJM and related Chern-Simons-matter theories, which recent references have studied \cite{Bagchi:2026emg, Bagchi:2024efs}. 

Our objective is to work solely from the bulk -- therefore we calculate bulk partition functions of black holes dual to ABJM, and calculate their flat space limit. However, several BPS black hole solutions dual to ABJM do not admit a flat space limit, most notably the magnetically charged black holes from \cite{Hristov:2013spa}. We will therefore identify a particular black hole solution -- the $U(1)^4$ BPS Kerr-Newman-$AdS_4$ solution -- which admits a non-trivial flat space limit in which the angular momentum is switched off. The geometry and the on-shell actions reduce to those of the BPS Reisner-Nördstrom black hole in flat space. We then reproduce the entropy of the latter by defining a suitable limit in the dual ABJM theory on $S^1\times S^2$, thus providing an instance of the construction in Figure \ref{Scheme} (at the level of the entropy). 
\\

\noindent\textbf{Outline. }Our paper is organised as follows: we start in section \ref{sec:AdSS} with the simplest example which are Schwarzschild-$AdS_4$ black holes, and study their on-shell actions and flat space limit, making explicit the contribution of small and large black holes in the limiting procedure. We then extend the discussion to charged black holes in minimal supergravity in section \ref{sec:charged}, studying the flat space limit of quantum corrections to their partition functions in the near-extremal regime in section \ref{sec:near_ext.}. We address rotating black holes in minimal supergravity in section \ref{sec:rotatingBH}, and finally charged black holes in gauged supergravity coupled to vector multiplets in section \ref{sec:gauged_sugra}, where  general remarks regarding the flat space limit of dyonic black holes with charge quantisation are made. In section \ref{sec:Carrol} we study the BPS Kerr-Newman-$AdS_4$ black hole in gauged supergravity, writing down its entropy in the flat space limit and reviewing some results which microscopically reproduce this entropy from the free energy of ABJM on $S^1\times S^2$. We then define a suitable limit on the ABJM charges and entropy function which allows us to microscopically reproduce the entropy of the BPS black hole in flat limit which is written earlier in this section.

\section{$AdS_4$-Schwarzschild black holes}\label{sec:AdSS}

We begin by exploring features of the semiclassical partition function in the flat limit for the simplest static, spherically symmetric black hole in $AdS_4$. We will highlight the distinction between the contributions of small and large black holes in this limit. 

\subsection{On-shell action and flat space limit}\label{on-shellSch}
We present the well-studied on-shell action for the Schwarzschild-AdS black hole. Recall the metric of Schwarzschild-AdS,
\begin{equation}
ds^2=-f(r)dt^2+\frac{dr^2}{f(r)}+r^2d\Omega^{2}_{d-2}
\end{equation}
where,
\begin{equation}
f(r)=\left(1-\frac{2M}{r}+\frac{r^2}{\ell_{AdS}^2}\right)
\end{equation}
The on-shell action is computed via standard holographic renormalisation, by adding suitable counterterms, in the spirit of \cite{Marino:2011nm}.
\begin{equation}\label{I_on-shell}
I_{\text{on-shell}}=I_{\text{bulk}}+I_{\text{surface}}+I_{\text{ct}}
\end{equation}
The Euclidean bulk Einstein-Hilbert action is given by,
\begin{equation}
I_{\text{bulk}}=\frac{-1}{16\pi G_N}\int_{M}d^{n+1}x\sqrt{G}\left(R-2\Lambda\right)
\end{equation}
where $G$ is in $(n+1)$-dims and $\Lambda$ is the cosmological constant. This term is divergent and will need to be regularised. The surface term is given by,
\begin{equation}
I_{\text{surface}}=\frac{-1}{8\pi G_N}\int_{\partial M}K|\gamma|^{1/2}d^nx
\end{equation}
where $\gamma$ is the metric induced by $G$ on the boundary.
The counter term action is obtained by regularising $I_{\text{bulk}}$. The insight from \cite{Emparan:1999pm,Balasubramanian:1999re} is that one may write a generic counterterm action using only combinations of curvature invariants in terms of the induced metric at the boundary. The counterterm action is built as an expansion in powers of the boundary curvature invariants, where the number of invariants needed to renormalise the bulk action grows with the spacetime dimension. This allowed \cite{Emparan:1999pm} to compute the on-shell action of asymptotically AdS geometries for coordinate systems corresponding to all possible foliations - spherical, flat or hyperbolic - with or without a thermal circle. 

We now turn to black holes in AdS with general $f(r)$. First Euclideanise the metric,
\begin{equation}\label{Euclidean_ds^2}
ds^2=+f(r)d\tau^2+\frac{dr^2}{f(r)}+r^2d\Omega^{2}_{d-2}
\end{equation}
where $t$ and $\tau$ are related by $\tau=it$. Then \cite{Marino:2011nm},
\begin{equation}
\begin{aligned}\label{action1}
I_{\text{bulk}}&=\frac{-1}{8\pi G_N}\frac{1}{\ell_{AdS}^2}r^n\text{vol}(S^{n-1})\beta\\
I_{\text{surface}}&=\frac{1}{8\pi G_N}\text{vol}(S^{n-1})\beta r^{n-1}\left(\frac{n-1}{r}f(r)+\frac{1}{2}f'(r)\right)\\
I_{\text{ct}}&=\frac{-1}{8\pi G_N}\text{vol}(S^{n-1}) r^{n-1}\left(\frac{n-1}{\ell_{AdS}}+\frac{(n-1)\ell_{AdS}}{2r^2}\right)\sqrt{f(r)}
\end{aligned}
\end{equation}
where $I_{\text{surface}}$ and $I_{\text{ct}}$ are understood to be evaluated at the boundary of AdS. 

First of all, for empty $AdS_4$ the on-shell action with the above scheme for Euclidean AdS with an $S^1\times S^2$ boundary vanishes - in agreement with the result found from \eqref{action1} for this geometry. In fact, this occurs for thermal $AdS_{n+1}$ with an $S^1\times S^{n-1}$ boundary for any odd $n$, as noted in \cite{Emparan:1999pm}.

When the geometry has a horizon, the total on-shell action equals the subtraction between the above expression evaluated at the boundary of AdS $r\to\infty$ and at the horizon, $I_\text{on-shell}=I|_{r=r_h}-I|_{r\to\infty}$. By carefully expanding the above expression in powers of $r^{-1}$, we see that the divergences localised close to the boundary in $I_{\text{surface}}$ are cancelled by those in $I_{\text{ct}}$, by construction.\footnote{Concretely, this is $\left(\frac{2}{\ell_{AdS}}+\frac{\ell_{AdS}}{r^2}\right)\left(1-\frac{2M}{r}+\frac{r^2}{\ell_{AdS}^{2}}\right)^{1/2}=\frac{2r}{\ell_{AdS}}+\frac{2}{r}-\frac{2M}{r^2}+\mathcal{O}(r^{-3})$.} Meanwhile, the gravitational Gibbons-Hawking term contributes with a finite piece in the limit where the boundary is pushed to infinity, namely $\lim_{r\to\infty}I=-\frac{\text{vol}(\mathbb{S}^2)\beta}{8\pi G_N} M$. The total result is
\begin{equation}\label{I_total}
    I^{\text{S-AdS}_4}_{\text{on-shell}}=\frac{\text{vol}(S^2)\beta}{8\pi G_N}\left(-\frac{r_h^3}{\ell_{AdS}^2}+ M\right)\text{.}
\end{equation}

Let us now take the large and small black hole limits, which are respectively $r_h/\ell_{AdS}\gg1$ and $r_h/\ell_{AdS}\ll1$. These are

\begin{align}\label{I_on-shell_SAdS}
I^{\text{S-AdS}_4}_{\text{on-shell}}\big|_{r_h/\ell_{AdS}\gg 1} &\;\longrightarrow\; \frac{-\beta}{2G_N}\frac{r_h^3}{\ell_{AdS}^2}  \\
I^{\text{S-AdS}_4}_{\text{on-shell}}\big|_{r_h\ll \ell_{AdS}} &\;\longrightarrow\; \frac{\beta}{2G_N}M
\end{align}

We may now use these expressions to write down the corresponding partition functions in the saddle-point approximation, which is the correct regime to study in the supergravity theory:
\be
Z\big|_{\text{sugra}}=e^{-I_{\text{on-shell}}}\text{.}
\ee
Consider now the flat space limit of the above expressions, corresponding to $\ell_{AdS}\to\infty$. Because big black holes have a horizon radius much larger than $\ell_{AdS}$, then this limit implies $r_{h,\text{big}}/\ell_{AdS}\to\infty$.\footnote{According to \eqref{r_h} the horizon radius of big black holes is $r_h=(2M)^{\frac{1}{3}}\ell_{AdS}^{\frac{2}{3}}$, which does not immediately imply that $r_h/\ell_{AdS}=\left(\frac{2M}{\ell_{AdS}}\right)^{\frac{1}{3}}\gg 1$ in the limit. However, we take this as the working definition of big AdS black holes; in other words, we require that their mass in AdS units is large so that the relation $r_h/\ell_{AdS}\to\infty$ is consistent as $\ell_{AdS}\to\infty$.} Therefore, the corresponding on-shell action in \eqref{I_on-shell_SAdS} diverges, leading to a diverging partition function for big black holes,
\be\label{DivZ}
\lim_{r_h/\ell_{AdS}\to\infty}Z_{\text{big}}\sim\lim_{r_h/\ell_{AdS}\to\infty}e^{\frac{\beta r_h}{2G_N}\frac{r_h^2}{\ell_{AdS}^2}}\;\to\; \infty\text{.}
\ee
In the flat space limit of small AdS black holes, we may take the limit $r_h/\ell_{AdS}\to 0$. This is because small black holes have a horizon radius smaller than $\ell_{AdS}$ and generically it will remain finite upon taking $\ell_{AdS}\to\infty$. Moreover, small SAdS black holes should become asymptotically flat Schwarzschild black holes in this limit, so that $r_h$ becomes the horizon radius of the Schwarzschild black hole, $r_h=2mG_N$, where $m\equiv M/G_N$ is the ADM mass written explicitly in Planck units. We arrive at
\be
\lim_{r_h/\ell_{AdS}\to 0}Z_{\text{small}}\sim e^{-\frac{\beta}{2}m}=e^{-S_{\text{Schwarzschild}}}\text{,}
\ee
where we make explicit that this result equals the saddle-point partition function for a Schwarzschild black hole, whose on-shell action equals minus its Bekenstein-Hawking entropy, $S_{\text{Schwarzschild}}=\frac{A}{4G_N}=4\pi G_Nm^2=\frac{M}{2T}$.

\subsection{Small vs large black holes}

From \eqref{DivZ}, observe that large S$AdS_4$ black holes give a diverging contribution to the on-shell semiclassical partition function in the flat-space limit. Consequently, the small S$AdS_4$ black holes give the \textit{finite} contribution to the partition function, precisely matching that of asymptotically flat Schwarzschild black holes when taking the limit. This agrees with large black holes increasingly dominating the canonical ensemble as their radius increases (their free energy increases in absolute value) and is pushed to null infinity in the flat space limit. Hence the only black hole solutions with finite action that survive the flat space limit are the ones arising from small AdS black holes.

Importantly, note that the divergence arising from large black holes is physical, and should be interpreted as an infinite volume divergence in a thermodynamic system. To see this, we write out explicitly the horizon radius in the large black hole limit in \eqref{I_on-shell_SAdS} as a function of the inverse temperature.\footnote{Using that the Hawking temperature for a SAdS large black hole is $T=\frac{1+\frac{3r^2_h}{\ell_{AdS}^2}}{4\pi r_h}$, which tends to $T\to\frac{3}{4\pi}\frac{r_h}{\ell_{AdS}^2}$ in the flat space limit.} Crucially we work in the canonical ensemble where $\beta$ is kept fixed. Therefore, to look carefully at how the on-shell action scales in the flat space limit, we should write all the quantities in terms of $\beta$ and $\ell_{AdS}$ which are then sent to infinity. We have
\begin{equation}
    \lim_{r_h/\ell_{AdS}\to\infty}\log Z_{ \text{big}} \sim \beta \frac{r^3_h}{2G_N\ell_{AdS}^2}=\left(\frac{4\pi}{3}\right)^3\frac{\ell_{AdS}^4}{2G_N\beta^2}\ \text{,}
\end{equation}
The final step is to rewrite the four-dimensional Newton's constant in terms of $\ell_{AdS}$ and the number of colors, or deconfined degrees of freedom, of the dual field theory, assuming a particular holographic duality. In four dimensions this will typically take the form $G_N\sim \ell_{AdS}^2/N^{\frac{3}{2}}$, \footnote{See for instance \cite{Marino:2011nm} for the precise relation in terms of the number of $M2$ branes and the volume of the $SE7$ manifold in the case of the duality between $AdS_4$ and vacuum $U(N)_k$ ABJM.} so that the above expression is, up to proportionality factors, $\sim \frac{\ell_{AdS}^2}{\beta^2}$. This is consistent with expectations from the putative dual field theory on $S_{\beta}^1\times S_{\ell_{AdS}}^2$, whose free energy in the infinite volume limit should have a volume divergence in the infinite volume limit. Because the radius of the $S^2$ is the same as that of AdS, this is precisely the limit which we are considering on the field theory side. More than that, if the field theory is conformal we expect its observables to depend on conformally invariant quantities, and since we have two length scales at our disposal the only such invariant quantity is $\ell_{AdS}/\beta$.

A final comment is that in the flat space limit this volume divergence should be present because the Hawking-Page transition temperature is pushed to zero for $\ell_{AdS}\to\infty$, so that the dominant saddle is always the large black hole. 

\section{Charged black holes in $4d$ minimal supergravity}\label{sec:charged}

In this section we consider charged, static and spherically symmetric black hole solutions of minimal supergravity. We extend the conclusions of the previous section by further distinguishing between non-extremal, extremal and BPS sectors of the partition function, and their relevant behaviour under the flat space limit.

\subsection{Geometry and properties}

The non-BPS asymptotically $AdS_4$ black holes have a macroscopic horizon and can be classified into large and small black holes. Much like neutral AdS black holes, the former have a horizon radius which scales with the AdS curvature radius and dominate the thermodynamic ensemble, while the latter are thermodynamically unstable. The blackening factor reads
\begin{equation}\label{AdS-RN_f(r)}
    f(r)=1-\frac{2M}{r}+\frac{Q^2}{r^2}+g^2r^2\text{.}
\end{equation}
Looking for its zeroes in the large horizon radius limit, we verify the above claim regarding large black holes, (using $g\propto \ell_{AdS}^{-1}$)
\begin{equation}
    \lim_{r_h/R\to\infty}f(r_h)=0\Rightarrow -\frac{2M}{r_h}+\frac{Q^2}{r_h^2}+\frac{r_h^2}{\ell_{AdS}^2}=0\text{.}
\end{equation}
Furthermore, $Q$ should not be arbitrarily large otherwise 1) the above equation has no roots and 2) the extremality bound is not satisfied. So assume $r_h^2/\ell_{AdS}^2\gg Q^2/r_h^2$ for very large black holes and we arrive at (this of course simply reproduces the analogous manipulation for the large horizon limit of the neutral AdS black hole)
\begin{equation}\label{r_h}
    \frac{r_h^3}{\ell_{AdS}^2}-2M=0\Rightarrow r_h=(2M\ell_{AdS}^2)^{\frac{1}{3}}\text{,}
\end{equation}
which indeed grows with $\ell_{AdS}$. 

Regarding the extremal bound of \eqref{AdS-RN_f(r)}, it is given by \cite{Chamblin:1999tk}
\be\label{extremality_bound}
3r_h^4+\ell^2_{AdS}r_h^2=Q^2\ell^2_{AdS}\text{.}
\ee
By taking the flat space limit of this relation $\ell_{AdS}\to\infty$ one recovers the flat space extremality condition. In contrast, the extremality bound in AdS is saturated at some value of $M$ which is greater than $Q$.

\subsection{On-shell action and flat limit}\label{charged_action}

\subsubsection{Non-BPS RN-AdS}

For a generic Lorentzian non-BPS solution we will take an electrically charged black hole; the profile of the gauge field takes the form \cite{Romans:1991nq}
\begin{equation}
F_{tr}=\frac{Q}{r^2}\text{.}
\end{equation}
We will work in the canonical ensemble where the gauge potential is allowed to fluctuate at the boundary, while the charge is held fixed and taken as an input in the statistical ensemble.
The on-shell action can be computed using various subtraction schemes to regularise.  In \cite{Chamblin:1999tk}, this is computed by subtracting the action of the extremal solution, and equals
\begin{equation}\label{I_Emparan}
    I^{RNAdS}_{\text{on-shell}}=\frac{\text{vol}(S^{2})\beta}{16\pi G_{N}\ell_{AdS}^2}\left(\ell_{AdS}^2 r_h-r_h^3+\frac{3Q^2\ell_{AdS}^2}{r_h}-\frac{4}{3}\ell_{AdS}^2r_e-\frac{8}{3}\frac{Q^2\ell_{AdS}^2}{r_e}\right)\text{,}
\end{equation}
This is done having in mind that the divergences in the on-shell action of AdS black holes are short-distance divergences originating from the near-boundary region, as discussed above. Since both geometries are asymptotically AdS such divergences will be cancelled, with additional finite terms which are contributions from the extremal horizon; these are the last two terms above. An important subtlety that we wish to point out is that fixing the period of the thermal circle to be equal to $\beta$ in the extremal solution is required for the asymptotic matching in this scheme to work. We will argue in the next subsection that there is an implicit order of limits in the assumption that the parameter $\beta$ in the extremal solution can be chosen to match the inverse temperature of a non-extremal solution to yield the expression \eqref{I_Emparan}.

Meanwhile, the renormalisation scheme employed when obtaining \eqref{I_total} simply subtracts boundary divergences with no reference to a background geometry which is subtracted. The bulk Maxwell action reads
\begin{equation}\label{maxwell_bulk}
    \frac{1}{16\pi G_N}\int\sqrt{g}F^{\mu\nu}F_{\mu\nu}=\frac{\text{vol}(S^2)\beta}{8\pi G_N}\int_{\infty}^{r_h}drr^2(F_{\tau r})^2=\frac{\text{vol}(S^2)\beta}{8\pi G_N}\frac{Q^2}{r_h}\text{,}
\end{equation}
At this point we have Euclideanised the metric and the components of the field strength; also $r_h$ denotes the outer horizon. The Maxwell term does not give any contributions to the counterterms \cite{Emparan:1999pm} and the Gibbons-Hawking term for the gauge field vanishes when the boundary of AdS is pushed to $r\to\infty$. Thus the total action is the above term summed with \eqref{I_total}, and its flat space limit for small black holes equals
\begin{equation}\label{I_RN_flat}
    \lim_{r_h/\ell_{AdS}\to 0}I^{RN\,AdS}_{\text{on-shell}}\to \frac{\beta}{2G_N}\left(M+\frac{Q^2}{r_h}\right)\text{,}
\end{equation}
where here $r_h=M+\sqrt{M^2-Q^2}$ is the outer horizon of the flat space RN black hole. This agrees with the result of \cite{Fairoos:2025kae}. On the other hand, when taking the large black hole limit the action diverges in the same way as the Schwarzschild-AdS studied in the previous section.

By rewriting $M$ in term of $r_h$ in the above expression we obtain the first three terms in \eqref{I_Emparan}, showing that the regularisation implemented in that result indeed corresponds to a subtraction of boundary divergences consistent with the holographic renormalisation procedure, supplemented with a subtraction of contributions from the on-shell action purely from the extremal horizon.

\subsubsection{Extremal limit of RN-AdS action and extremal entropy}
Recall the action for RN-AdS in 4d in the extremal subtraction scheme from \cite{Chamblin:1999tk},
\begin{equation}
    I^{RN-AdS}_{\text{on-shell}}=\frac{\text{vol}(S^2)\beta}{16\pi G_N}\left(r_h-\frac{r_h^3}{\ell_{AdS}^2}+\frac{3Q^2}{r_h}-\frac{4}{3}r_e-\frac{8}{3}\frac{Q^2}{r_e}\right)\,\text{,}
\end{equation}
where $r_e$ refers to the horizon radius of the extremal solution. As we have stated earlier, the validity of this expression requires that the periodicity of the thermal circle in the extremal background is allowed to be arbitrary, and then match it with the inverse temperature $\beta$ of the non-extremal solution whose action we wish to compute. This is the matching between the extremal background and the black hole background mentioned in \cite{Chamblin:1999tk} which allows for the asymptotic geometries of both background to be equal, thus cancelling boundary divergences.

A consequence of this is that the on-shell action of the extremal black hole to zero. This is in line with the result from \cite{Mitra:1998tv}, where it is the extremal entropy which vanishes in the fixed potential ensemble.\footnote{This is not a contradiction since the entropy is related to the on-shell action through a Legendre transform, which changes the ensemble under consideration.}

For the above to hold, it is important that the quantity $1+\frac{3r_e^2}{\ell_{AdS}^2}-\frac{Q^2}{r_e^2}$ is \textit{not} identified with ($4\pi r_e$ times) the temperature \cite{Mitra:1998tv}. One should instead leave the periodicity of the thermal circle $\beta$ as a temperature-independent constant, as in \cite{Chamblin:1999tk}.
Otherwise, the transition from Euclidean non-extremal geometries to extremal ones is non-smooth in the Gibbons-Hawking procedure \cite{Fairoos:2025kae}.

Computing the gravitational partition function of the extremal RN black hole in this way may be viewed as a procedure of "quantisation after extremisation" \cite{Mitra:1998tv,Ghosh:1996gp}, whereby the gravitational path integral, in the Gibbons-Hawking sense, is defined in an ensemble where $Q$ and $M$ are fixed and its evaluation is determined by the unique saddle satisfying the extremality bound, that is, the extremal RN-AdS black hole.

Moreover, as argued in \cite{Fairoos:2025kae}, the near-horizon geometry of an extremal black hole is of the form $AdS_2\times S^2$, which cannot be put in the form $dR^2+R^2d\theta^2+...$, making it impossible to identify a thermal circle in the first place. This argument applies to an order of limits where extremisation is performed at the beginning of the calculation, which is the relevant one to the cases discussed so far. This also explains why they have found a vanishing Euler characteristic for the extremal background, thus rendering $\beta$ ill-defined. We take the perspective that this argument actually shows that the thermal circle still exists but its period is not fixed by any regularity condition; such a condition cannot exist since the horizon is at an infinite proper distance away from any point at a finite coordinate radius in the exterior \cite{Hawking:1994ii,Carroll:2009maa}.

However we argue that the more appropriate order of limits is to perform extremisation at the end of the calculation; this means that 1) we do not impose the extremality bound and 2) we identify $\beta$ with the inverse temperature determined from $f'(r_h)$ which is left implicit. The extremal action will then be correctly identified with minus the Bekenstein-Hawking entropy when the appropriate limit on the extremality bound and $\beta$ is taken simultaneously.

Let us see how this works for the case of the flat space limit of small black holes. In this regime the action reduces to reduces to 
\begin{equation}\label{flat_Emparan}
      I^{RN-AdS}_{\text{on-shell}} \xrightarrow[]{\text{small}}\frac{\pi r_{h}^3}{r_{h}^2-Q^2}\left(\frac{3Q^2}{r_h}+r_h-4Q\right)\text{,}
\end{equation}
where we have used that \cite{Chamblin:1999tk}
\begin{equation}\label{beta}
   \beta=\frac{4\pi\ell_{AdS}^2r_{h}^3}{3r_{h}^4+\ell_{AdS}^2r_{h}^2-Q^2\ell_{AdS}^2}\text{.}
\end{equation}

Let us first check that, in the non-extremal regime, this on-shell action reproduces the Bekenstein-Hawking entropy. In the canonical ensemble (fixed temperature) the entropy is given by
\begin{equation}\label{entropy}
    S=\beta\frac{d }{d \beta}I_{\text{on-shell}}-I_{\text{on-shell}}\text{.}
\end{equation}
There are two types of contributions to this expression. The on-shell actions always contain an overall factor of $\beta$ from the Euclidean time integral, and the differentiation of that factor in the first term always cancels when the second term is subtracted. Importantly, the contribution from the finite piece in \eqref{I_Emparan} will drop out in this subtraction\footnote{In the fixed charge ensemble, $r_e$ is fixed in terms of $Q$ and is independent of $\beta$.} and the entropy will therefore equal that computed with \eqref{I_RN_flat}, making it clear that \eqref{entropy} is scheme-independent. To explicitly see the equality between the two schemes, one should remember that in the canonical ensemble $M$ in \eqref{I_RN_flat} should be expressed in terms of the outer horizon radius. The final step is to note that the non-zero contribution to the entropy comes from expressing $r_h$ in terms of $\beta$ through \eqref{beta}. In the end we obtain
\begin{equation}
    \begin{aligned}
        S&=\frac{\beta^2}{4}\frac{\partial r_h}{\partial\beta}\frac{\partial}{\partial r_h}\left(\frac{3Q^2}{r_h}+r_h\right)=\pi r_h^2=S_{BH}^{RN}\text{.}
    \end{aligned}
\end{equation}

If one decides to compute the extremal entropy by taking the limit $r_h\to Q$ of the entropy, one obtains a finite result equal to the Bekenstein-Hawking entropy. A similar limiting procedure can be taken at the level of the on-shell action, but there one faces the fact that $\beta$ is formally divergent in the extremal limit, so this has to be taken carefully. Take $r_h=q+\epsilon$, where $\epsilon\ll1$. This gives
\begin{equation}\label{eq:IRN_ext_flat}
       I^{RN-AdS}_{\text{on-shell}} \xrightarrow[2) \text{ extremal}]{1) \text{ small}} -\pi Q^2=-S_{BH}^{RN}\text{,}
\end{equation}
where we have noted that $r_e=Q$ is the extremal radius of the RN black hole in flat space, $S_{BH}^{RN}$ being its Bekenstein-Hawking entropy.\footnote{See also \cite{Page:2000dk,Hawking:1994ii,Azzurli:2017kxo}. In particular, in \cite{Azzurli:2017kxo} the entropy of non-extremal deformations of the extremal solution are computed so that the extremal entropy becomes well defined.}

The limit taken above is in the spirit of the procedure carried out in \cite{Carroll:2009maa} to compute the proper distance between the inner and outer horizons in the extremal limit, showing that it remains finite. However their conclusion that the semiclassical entropy of extremal black holes as captured by microscopic state counting from string theory should be associated to the compactified $AdS_2\times S^2$ saddle of the Euclidean path integral should not be modified by our arguments above, which concern the extremal limit of quantities computed for non extremal geometries. For instance in section 3 of \cite{Sen:2008vm} where it is also emphasised that Euclidean path integral methods cannot be applied directly to extremal geometries and one should take an appropriate limit of non-extremal Euclidean results.

\subsubsection{BPS charged $AdS_4$ black holes}

We now consider asymptotically $AdS_4$ BPS black hole solutions of $\mathcal{N}=2$ minimal gauged supergravity with a spherical horizon. These come in two classes \cite{Romans:1991nq}, the electric and magnetically charged solutions, whose blackening factor is given by (here and in the rest of this paper, we identify the gauging parameter as $g\equiv \ell_{AdS}^{-1}$)
\begin{equation}\label{BPS_sols}
    \begin{aligned}
    &\text{Electric:  }\,\,\,\,\,\,f(r)=\left(1-\frac{M}{r}\right)^2+g^2r^2\text{,}\\
    &\text{Magnetic: }\,\,\,\,\,f(r)=\left(gr+\frac{1}{2gr}\right)^2+\frac{M^2}{r^2}\,\text{.}
    \end{aligned}
\end{equation}
Both these solutions represent naked singularities. At level of the metric, the electric solution becomes the extremal RN black hole in flat space, while the magnetic one does not admit a flat space limit ($f(r)$ blows up as $g\to 0$). We will see in section \ref{sec:gauged_sugra} that this is a more general phenomena in magnetically charged black holes in gauged supergravity.

We can make use of the expressions presented so far to compute the on-shell action of BPS black holes \eqref{BPS_sols}. Since these black holes are naked singularities, they have zero horizon radius. Therefore our expression for the renormalised on-shell action is infinite, with the divergence coming from the bulk gauge field action \eqref{maxwell_bulk}.\footnote{One can still check that for both BPS solutions the expression \eqref{action1} removes all divergences at $r\to\infty$, leaving no finite piece. This calculation also presents a similarity with \cite{Halmagyi:2017hmw}, whose BPS on-shell actions also only receive contributions from the horizon and the finite pieces from the horizon vanish when evaluated at the BPS configurations} One can make sense of this divergence by considering the contribution of these solutions to the semiclassical partition function.

The partition function evaluated for each of these BPS on-shell geometries is
\begin{equation}
    Z_{\text{Electric}}=Z_{\text{Magnetic}}=\lim_{r_h\to 0}e^{-\frac{\text{vol}(\mathbb{S}^2)\beta}{8\pi G_N}\left(-\frac{r_h^2}{\ell_{AdS}^2}+\frac{Q^2}{r_h}\right)}=0\text{.}
\end{equation}
This shows that the $AdS_4$ BPS solutions of minimal $\mathcal{N}=2$ supergravity, being unphysical naked singularities, do not contribute at the level of semiclassical state-counting. 

There exist other magnetically charged BPS black holes in $AdS_4$ in minimal $\mathcal{N}=2$ supergravity, whose horizon topology is a Riemann surface of genus greater than $1$ \cite{Caldarelli:1998hg}. These were shown to have a finite on-shell action, for instance in \cite{Azzurli:2017kxo}. However, the metric of these solutions still blows up in the flat limit. This is a manifestation of the fact that black holes in asymptotically flat space can only have spherical horizons \cite{Hawking:1971vc}. We therefore consider these black holes to not contribute to the partition function in the flat limit.

To conclude the discussion on the Reissner-Nördstrom black holes we make a final comment based on the two conclusions from this section:
\begin{enumerate}
    \item The non-BPS $AdS_4$ black holes have an on-shell action which reproduces that of the flat space RN $AdS_4$ BH upon taking the flat space limit; if one starts from the extremal AdS-RN, its on-shell action will equal minus the Bekenstein-Hawking entropy of the extremal BPS RN in flat space in the limit;
    \item The BPS $AdS_4$ geometries, having naked singularities, have a positively diverging on-shell action and their partition function vanishes before taking the flat space limit.
\end{enumerate}
These two facts imply that the contribution to the partition function from BPS-RN in flat space arises entirely from flat space limit of the non-BPS AdS solution. Notice that the metric of the extremal RN flat space solution can be seen to arise as the flat limit of either the BPS $AdS_4$ electric solution shown above or the extremal $AdS_4$ RN solution. This subsection clarifies that the partition function is inherited uniquely from that of the flat limit of the extremal $AdS_4$ solution. 

We also note that point 1. above implies an enhancement of preserved supercharges of states in the flat space limit, given that the contribution to the partition function of a $\frac{1}{2}$-BPS geometry originates from that of a non-BPS geometry.

\section{Near-extremal black holes}\label{sec:near_ext.}

We now extend our analysis to near-extremal black holes and study quantum corrections to the flat space  partition function in the low temperature regime. The corrections are subleading in the canonical ensemble. To this end, we study the near-extremal Reissner-Nördstrom black hole and proceed along the lines of \cite{Iliesiu:2020qvm}. In this work, they dimensionally reduced the near-extremal geometry to obtain JT-gravity with a dual Schwarzian theory. This setup accurately captures the corrections to the partition function, and a further analysis also can be done to investigate the mass gap via computing the density of states, which will be commented on further.

\subsection{Setup}

Starting from the RN black hole in $AdS_4$ studied in the previous section, the near-extremal geometry is obtained by writing the horizon radius as $r_h=r_0+\delta r_h$, with $\delta r_h\ll r_0$. Here $r_0$ is the extremal radius and the double pole of $f(r)$ in the extremal limit, while $\delta r_h$ can be expressed in terms of the inverse temperature $\beta$ as
\begin{equation}
    \delta r_h=\frac{2\pi}{\beta}\frac{\ell_{AdS}^2r_0^2}{\ell_{AdS}^2+6r_0^2}+...
\end{equation}
 Here the ellipses represent subleading terms in the low temperature regime. The extremal mass, charge and Bekenstein-Hawking entropy in terms of $r_0$ are given by, 
\begin{align}
    Q^2=& \frac{4\pi}{G_N}\left(r_0^2+\frac{3r_0^4}{\ell_{AdS}^2}\right)\,, \,\,\,\,\,\,\,\,\,\,\,\,\,
    M_0=\frac{r_0}{G_N}\left(1+\frac{2r_0^2}{\ell_{AdS}^2}\right)\, ,\,\,\,\,\,\,\,\,\,\,\,\,\,\,\,\,    S_0=\frac{\pi r_0^2}{G_N}
\end{align}

In the near extremal limit, the bulk geometry is separated into to sub regions: the near horizon and far away region. The former is approximately $AdS_2\times S^2$. The sphere radius is set by $r_0$. In the far away region, the geometry is approximately extremal $AdS_4$. The $\text{AdS}_2$ radius is given by  $\ell_2=\frac{r_0\ell_{AdS}}{\sqrt{\ell^2+6r_0^2}}$, which sets the scale wherein quantum corrections at the $\text{AdS}_2$ throat become relevant. In particular, writing the mass and entropy of the near-extremal black hole as functions of $\beta$ and $Q$, the temperature corrections are suppressed by the gap scale $M_{\text{SL}(2)}^{-1}\equiv \frac{r_0\ell_2^2}{G_N}$.

\subsection{Partition function}
The quantum corrections to the partition function are computed by dimensionally reducing the theory near the horizon to a $2d$ Jackiw-Teitelboim (JT) gravity theory. The theory is reduced on $S^2$ in the near horizon region and the resulting geometry is the hyperbolic disk. The boundary Schwarzian theory is then the path integral on this disk. In the fixed electric charge ensemble, the partition function of the near-extremal RN black hole in $AdS_4$ in the fixed charge ensemble is computed by evaluating,
\begin{equation}\label{Schwarz}
    Z_{RN}=e^{\pi\Phi_0(Q)}e^{-\beta M_0(Q)}\int\frac{\mathcal{D}\mu[F]}{SL(2,\mathbb{R})}e^{\pi\Phi_{b,Q}\int_0^\beta du \text{Sch}(F,u)}\,.
\end{equation}
Here, $\Phi_0(Q)$ is the (charge dependent) value of the dilaton at the horizon, $M_0(Q)$ is the extremal mass, $\Phi_{b,Q}$ relates to the value of the dilaton at the boundary of the near-horizon throat.\footnote{See \cite{Iliesiu:2020qvm} for further details.} The measure of the partition function is given by the Weil-Petersson symplectic form.\footnote{Details on how to obtain the measure and solve this path integral are given in \cite{Mertens:2022irh} and related references.} The Schwarzian is given by,
\begin{equation}
    \operatorname{Sch}(F, u)=\frac{F^{\prime \prime \prime}}{F^{\prime}}-\frac{3}{2}\left(\frac{F^{\prime \prime}}{F^{\prime}}\right)^2
\end{equation}
Where, $F(u)=\tan\left(\frac{\pi\tau(u)}{\beta}\right)$ and $\tau(u+\beta)=\tau(u)+\beta$ are used to transform Poincaré coordinates to the disk.

The first exponential term in \eqref{Schwarz} is the area contribution from the dilaton action in the near horizon regime. The second piece comes from evaluating the on-shell action in the far away region. The fluctuations from the dilaton are captured by the Schwarzian path integral. This is essentially the disk partition function, and can be computed (one-loop) exactly,
\begin{equation}\label{discpartitionfunct}
     Z_{RN}=\left(\frac{\Phi_{b,Q}}{\beta}\right)^{3/2}e^{\pi\Phi_0(Q)}e^{-\beta M_0(q)}e^{\frac{2\pi^2}{\beta}\Phi_{b,Q_0}}\ .
\end{equation}
\\
The first piece comes from the one-loop determinant. This gives a logarithmic correction to the free energy. The second and third term respectively are the extremal entropy and extremal mass term. The fourth piece comes from the semiclassical near-extremal contribution. The corresponding entropy is,
\begin{equation}
    S(\beta,Q)=S_0+\frac{4\pi^2\Phi_{b,Q}}{\beta}-\frac{3}{2}\log\frac{\beta}{e\Phi_{b,Q}}\ .
\end{equation}
 $\beta M_0$ corresponds to an expansion of the semiclassical action around extremal result.

\subsection{Flat space limit}

We now examine the flat space limit of the near-extremal partition function that we have reviewed in the previous subsection. The flat space limit of \eqref{discpartitionfunct}, obtained from $\ell_{AdS}\to\infty$, is
\begin{equation}\label{Flatdiscpartition}
\begin{aligned}
    Z_{RN}&=\left(\frac{r_0\ell_2^2}{\beta}\right)^{3/2}\text{exp}\left[\frac{\pi r_0^2}{G_N}-\beta M_0(Q)+\frac{2\pi^2}{\beta}\frac{r_0\ell_2^2}{G_N}\right]\\
    \overset{\text{flat}}\longrightarrow &\left(\frac{r_0^3}{\beta}\right)^{3/2}\text{exp}\left[\frac{\pi r_0^2}{G_N}+\frac{2\pi^2}{\beta}\frac{r_0^3}{G_N}\right]\,,
\end{aligned}
\end{equation}
where we have used the definitions in \cite{Iliesiu:2020qvm},
\begin{equation}\label{definitions}
\begin{aligned}
  M_0&=\frac{2r_0^3}{G_N\ell_{AdS}^2}\,,\quad\ell_2=\frac{\ell_{AdS} r_0}{\sqrt{\ell_{AdS}^2+6r_0^2}}\,,\\
  \Phi_{b,Q}&=\frac{r_0\ell_2^2}{G_N}\,,\quad \Phi_0=\frac{r_0^2}{G_N}\,. 
\end{aligned}
\end{equation}
The expression \eqref{Flatdiscpartition} can be interpreted as a one-loop correction to a flat space partition function in JT gravity. The extremal mass contribution drops out, the area term remains and the semi-classical correction from the (now flat space) near-extremal contribution stays. 

We can also make contact with \cite{Heydeman:2020hhw}, which computes the partition function of near-BPS black holes in flat space. In order to compare with the flat limit of the partition function of the AdS black hole shown above, we should discard the contributions from the one-loops fermionic determinant and from the SU$(2)$ modes in \cite{Heydeman:2020hhw}, and retain contributions from the purely bosonic Schwarzian theory,
\begin{equation}
    Z=\left(\frac{\Phi_r}{\beta}\right)^{\frac{3}{2}}\exp\left[\frac{2\pi^2}{\beta}\Phi_r\right].
\end{equation}

Here $\Phi_r$ is the boundary condition for the dilaton in the near-horizon throat of the asymptotically flat geometry, which takes the form
\begin{equation}\label{Z_flat}
    \Phi_r=\frac{r_0\ell_2^2}{G_N}=\frac{r_0^3}{G_N}\,,
\end{equation}
where we used that for the asymptotically flat geometry, $\ell_2=r_0$. This of course equals the $\ell_{AdS}\to\infty$ limit of the boundary condition for the dilaton in the asymptotically AdS geometry $\Phi_{b,Q}$ in \eqref{definitions}. From \eqref{Z_flat} we also see that a term corresponding to the extremal mass in \eqref{discpartitionfunct}, which is a contribution from the FAR region in \cite{Iliesiu:2020qvm}, should be absent; this is consistent with the limit taken in \eqref{Flatdiscpartition}. The remaining pieces in \eqref{Z_flat}, namely the Schwarzian one-loop determinant and the expansion of the semiclassical extremal piece remain, and equal those in \eqref{Flatdiscpartition} when taking the flat limit from the AdS partition function. We then conclude that the one-loop corrections to the partition function of charged near-extremal black holes in flat space come entirely from the one-loop corrections of the flat limit of their AdS counterparts.

One of the objectives of \cite{Iliesiu:2020qvm} was to show the existence of the mass gap by computing the density of states. It was shown that there is no mass gap due to the continuous nature of the density of states. A consequence of our analysis is that this fact survives also in the flat space limit.

 One can also consider one-loop corrections to the partition function of extremal black holes. The corrections to  black hole entropy computed in \cite{Banerjee:2010qc,Banerjee:2011jp,Sen:2012kpz} are applicable only to BPS and extremal black holes and are therefore temperature independent.\footnote{These are logarithmic in the area, and can be obtained by integrating the zero mode of the heat kernel associated to the mode expansion of massless supergravity fields.} In fact, such logarithmic corrections can be recovered from the construction of \cite{Iliesiu:2020qvm,Heydeman:2020hhw} in the form of massive KK modes originating in the dimensional reduction of supergravity fields from 4D to 2D. These give corrections to the extremal, semiclassical entropy $S_0$ in the form $\delta S_0\propto \log r_0$, with the proportionality factor depending on the 4D matter content.

 We also would like to clarify that the near-extremal solution in the flat limit produces the usual $AdS_2\times S^2$ near-horizon geometry, therefore the effective theory governing the near-extremal regime is still a theory of JT gravity whose on-shell solutions are negatively curved. This is not the same as flat JT gravity, which would describe near-extremal black holes with an $\mathbb{R}^{1,1}\times S^2$ near-horizon geometry.

\section{Rotating black holes in minimal supergravity}\label{sec:rotatingBH}
We consider a charged rotating black hole in AdS with metric in Boyer-Lindquist type coordinates,
\begin{equation}
d s^2=-\frac{\Delta_r}{\rho^2}\left[d t-\frac{a \sin ^2 \theta}{\Xi} d \phi\right]^2+\frac{\rho^2}{\Delta_r} d r^2+\frac{\rho^2}{\Delta_\theta} d \theta^2+\frac{\Delta_\theta \sin ^2 \theta}{\rho^2}\left[a d t-\frac{r^2+a^2}{\Xi} d \phi\right]^2\,,
\end{equation}
where
\begin{equation}\label{paramskerr}
\begin{aligned}
&\rho^2=r^2+a^2 \cos ^2 \theta, \quad \Xi=1-\frac{a^2}{\ell_{AdS}^2} \\
&\Delta_r=\left(r^2+a^2\right)\left(1+\frac{r^2}{\ell_{AdS}^2}\right)-2 m r+z^2, \quad \Delta_\theta=1-\frac{a^2}{\ell_{AdS}^2} \cos ^2 \theta \,.
\end{aligned}
\end{equation}
Here, $a$ is the rotational parameter and $z$ is a parameter defined by $z^2=q_e^2+q_m^2$, the electric and magnetic charges.

\subsection{Supersymmetry and the BPS limit}\label{susylimit}
Here we recollect the conditions for which an amount of the supersymmetry is preserved. For the following conditions we have that one quarter of the supersymmetries are preserved \cite{Caldarelli:1998hg,Alonso-Alberca:2000zeh},
\begin{equation}
    m^2=a\,\ell_{AdS} \left(1+\frac{a}{\ell_{AdS}}\right)^4, \quad q_e^2=a\,\ell_{AdS} \left(1+\frac{a}{\ell_{AdS}}\right)^2, \quad q_m=0
\end{equation}
In the BPS limit we have,
\begin{equation}\label{charges_BPS_Kerr}
    M=\frac{\sqrt{a\,\ell_{AdS} }}{\left(1-\frac{a}{\ell_{AdS}}\right)^2}, \quad Q_e=\frac{\sqrt{a\,\ell_{AdS} }}{1-\frac{a}{\ell_{AdS}}}, \quad Q_m=0, \quad J=\frac{\sqrt{a\,\ell_{AdS} }a }{\left(1-\frac{a}{\ell_{AdS}}\right)^2}
\end{equation}
and we recover the BPS bound
\begin{equation}\label{boundKerr}
    M=Q_e+\frac{J}{\ell_{AdS}}
\end{equation}
We also note the horizon radius and the angular velocity,
\begin{equation}\label{r+}
    r_h^2=a\ell_{AdS}\,,\,\,\,\,\,\,\,\,\,\,\,\,\,\Omega=\frac{1}{\ell_{AdS}}
\end{equation}
As remarked in \cite{Caldarelli:1999xj}, this means that the CFT dual to this BPS black hole is living on a background with an angular velocity equal to the speed of light.

\subsection{On-Shell Action}

The on-shell action is calculated familiarly, employing the scheme of \cite{Emparan:1999pm}, by adding a suitable $I_{ct}$ to regularise the diverging integrals. With this, the Euclidean action for the non-extremal solution is straightforwardly computed to be,
\begin{equation}\label{KerrAction}
I^{\text{Kerr}}_{\text{on-shell}}=\frac{\beta}{4 G \ell_{AdS}^2 \Xi}\left[-r_{+}^3+\Xi \ell_{AdS}^2 r_{+}+\frac{\ell_{AdS}^2\left(a^2+z^2\right)}{r_{+}}+2 \frac{\ell_{AdS}^2 z^2 r_{+}}{r_{+}^2+a^2}\right],
\end{equation}
where $z^2\equiv q_e^2+q_m^2$ and the inverse temperature is
\begin{equation}
\beta=\frac{4 \left(a^2+r_h^2\right)}{r_h \left(-\frac{a^2+z^2}{r_h^2}+\frac{a^2}{\ell_{AdS}^2}+\frac{3 r_h^2}{\ell_{AdS}^2}+1\right)}\,.
 \end{equation}
 
\subsection{Flat Limit and a suitable scaling in the BPS limit}\label{flat limit Kerr}
Let us now take the flat space limit by sending $\ell_{AdS}\rightarrow \infty$. The inverse temperature in this limit becomes,
\begin{equation}
    \beta \overset{\ell_{AdS}\rightarrow \infty}{=}\frac{4 a^2 r_h+4 r_h^3}{-a^2-z^2+r_h^2}\,.
\end{equation}
The on-shell action of the non-extremal solution becomes
\begin{equation}\label{eq:IKerr_flat}
    \tilde{I}_{\text{flat}}=-\frac{1}{G}\frac{\left(a^2+3 z^2\right) r_h^2+a^2 \left(a^2+z^2\right)}{a^2+z^2-r_h^2}\,.
\end{equation}
Let us analyse this expression in several cases:
\begin{itemize}

\item First of all, setting $a=0$ after taking the flat limit of the Kerr-AdS black hole produces the metric and on-shell action of the flat limit of the RN black hole in flat space, as would be expected (taking into account that this expression does not include the extremal subtraction present in \eqref{I_Emparan}). 

\item If we instead consider taking the $\ell_{AdS}\to\infty$ limit by setting the parameters of the solution to their BPS values from section \ref{susylimit}, we see that the on-shell action is divergent. This is because the horizon radius of the Kerr-AdS black hole in the BPS limit scales with the AdS radius, see \eqref{r+}. Similarly, all of its charges \eqref{charges_BPS_Kerr} also blow up, being related by the BPS condition \eqref{boundKerr}.

\item However, there \textit{is} a well-defined flat space limit of the Kerr-AdS black hole, both at the level of the metric and the on-shell action. This is
\begin{tcolorbox}  
\begin{equation}\label{appropriate_limit}  \ell_{AdS}\to\infty\,,\,\,\,\,\,\,\,\,\,a\to 0\,,\,\,\,\,\,\,\,\, a\,\ell_{AdS}=\text{constant.}
\end{equation}
\end{tcolorbox}
In this limit, the angular momentum vanishes and we obtain an extremal (and BPS) RN black hole in flat space, with $M=Q=z$, where $z^2\equiv a\,\ell_{AdS}$. Its on-shell action is the one analysed in section \ref{charged_action}, namely \eqref{flat_Emparan}, with $Q$ given by the square root of the constant defined in the limit above \eqref{appropriate_limit}. Not only that, the horizon radius remains finite. This allows the particular limit \eqref{appropriate_limit} of the Kerr-AdS black hole to contribute finitely to the partition function in the flat limit (in the BPS sector). Note that this represents an enhancement of supersymmetry in this sector of the semiclassical partition function from $\frac{1}{4}$-BPS to $\frac{1}{2}$-BPS.

\item One can check that taking this limit on the non-extremal Kerr-AdS black hole also produces the metric and the on-shell action of a non-extremal RN black hole.
\end{itemize}

We emphasise that this is the only black hole solution in the minimal sector of $4D$ $\mathcal{N}=2$ gauged supergravity where a regular flat limit can be defined, in the sense that BPS black holes with a spherical horizon are generically naked singularities when rotation is absent.\footnote{See for instance Table 1 of \cite{Hristov:2013spa} for an overview of known BPS solutions according to their horizon topology, preserved supersymmetries and absence/ presence of a horizon.} These are therefore not desirable states to consider in the flat space limit. In the next section we will discuss the conditions under which black holes in gauged supergravity coupled to vector multiplets can admit a flat limit.

\section{Charged black holes in non-minimal gauged supergravity}\label{sec:gauged_sugra}

One of the main takeaways of the previous section is that, regarding $4D$ $\mathcal{N}=2$ minimal supergravity, the mapping between black hole states in flat space and with a negative cosmological constant is not bijective. In particular, the metric of the extremal and BPS charged black hole in flat space can arise from the flat limit of either the electric BPS solution of \cite{Romans:1991nq} or from the extremal RN in $AdS_4$. A precise calculation of on-shell actions was needed to understand this mapping at the level of semiclassical state-counting. 

With a view of making contact with supergravity solutions relevant for comparisons with exact computations of partition functions in 3D CFT's \cite{Benini:2015eyy}, we wish to extend our understanding of this mapping in the context of gauged non-minimal supergravity. 

Here we will consider solutions of the bosonic Lagrangian of $U(1)$ Fayet-Iliopoulos gauged supergravity with $n_V$ vector multiplets and in the absence of hypermultiplets:
\begin{equation}
\mathcal{L}=\frac{1}{2}R+g_{i\bar{l}}\partial_\mu z^i\partial^\mu \bar{z}^{\bar{l}}+I_{\Lambda\Sigma}F^{\Lambda}_{\mu\nu}F^\Lambda_{\Sigma \mu\nu}+\frac{1}{2}R_{\Lambda\Sigma}\epsilon^{\mu\nu\rho\sigma}F_{\mu\nu}^\Lambda F^\Sigma_{\rho\sigma}-V_g\,,
\end{equation}
where $\Lambda,\Sigma=0,1,...,n_V$ and $i=1,...,n_V$, $R_{\Lambda\Sigma}$ and $I_{\Lambda\Sigma}$ are the real and imaginary parts of the period matrix $\mathcal{N}_{\Lambda\Sigma}$. The gauge fields couple to the gravitini through a linear combination of the graviphoton
and the $n_V$ vectors from the vector multiplets, $\xi_\Lambda A^\Lambda_\mu$. The constants $\xi_\Lambda$
are called FI parameters. The electric charges of the gravitinos are then denoted by $e_\Lambda\equiv g\xi_\Lambda$. Here we follow the conventions of \cite{Toldo:2012ec,Hristov:2010ri} for the quantities appearing above. 

\subsection{Quantisation condition $=$ Singular flat limit}\label{sec:quantisation}
 Here we bring to light aspects of the Dirac quantisation (or Dirac-Schwinger-Zwanziger quantisation) and its consequence in taking a flat space limit on dyonic black holes in gauged supergravity. This is a quantisation condition on the charges $p_\Lambda g^\Lambda=n,\quad n\in \mathbb{Z}$.\footnote{In general, the Dirac quantisation condition concerns a theory with an electric charge and a magnetic charge (in the context of supergravity, we mean electric and magnetic with respect to a fixed gauging $g_\Lambda$ \cite{Trigiante:2016mnt,Romans:1991nq}:
\[
p_2^\Lambda e_{1\,\Lambda}-p_1^\Lambda e_{2\,\Lambda}\in\mathbb{Z}
\]
In gauged supergravity, the gravitini are electrically charged under the vector potential of the gravity multiplet. This can therefore be thought of as the second charged particle in the above relation, with $e_2^\Lambda\propto g^\Lambda$ and $p_2^\Lambda=0$,
} which ensures that the gravitino are single valued when parallel transported along a closed cycle, and is unavoidable for consistency of the theory. For BPS solutions $n$ is usually restricted to be $\pm 1$ \cite{Hristov:2010ri}. This condition has also been imposed in \cite{Toldo:2012ec} for magnetic, non-extremal black hole solutions, in order that they asymptote to a supersymmetric, magnetically charged AdS vacuum.

The problem with enforcing the quantisation condition is clear when taking the flat space limit - it causes the magnetic charges to blow up: 
\begin{equation}
    p^\Lambda\propto\frac{1}{g}\;\;\Rightarrow \;\;p^\Lambda\underset{g\to 0}{\to}\infty\,.
\end{equation}
This in turn causes several quantities to diverge in the limit, namely the blackening factor of the black hole ansatz and the horizon radius.

Thus there appears to be a distinction between physical black hole states in AdS with regards to taking the flat limit: electric ones do not \textit{a priori} have any obstruction to this limit, while this is hindered in the magnetic ones due to the Dirac quantisation condition. This can be seen as a consequence of the failure of electric-magnetic duality, with respect to the same gauging, in the presence of a gauging potential: as emphasised in \cite{Gnecchi:2014cqa} the potential is invariant under a symplectic rotation of the charge matrix, and so are the solutions of the equations of motion provided that the gauging changes from electric to magnetic and vice versa. In particular, under the action of a symplectic transformation $S\in Sp(4,\mathbb{R})$, the gauging potential transforms as
\begin{equation}
    V_g(S\mathcal{V},\mathcal{G})=V_g(\mathcal{V},S\mathcal{G})\,,
\end{equation}
while the black hole potential is straightforwardly invariant. This ensures that a solution of the equations of motion is invariant under a symplectic transformation provided that electric charges are traded with magnetic charges, and the electric gauging defined by $g_\Lambda$ becomes a magnetic gauging defined by $\hat{g}^\Lambda=-\mathcal{I}^{\Lambda\Sigma}g_\Sigma$. However in the previous discussion we always refer to magnetic/ electric black holes \textit{with respect to the same gauging} $g_\Lambda$.
In the flat space limit, the gauging is removed and the gravitino is no longer electrically charged. Therefore, if they are seen as arising from the flat limit of AdS black holes, they should arise from solutions on which the quantisation condition is not imposed. These are usually electrically charged black holes.

Our conclusion is therefore that any excitation above the supersymmetric magnetic AdS vacuum, namely magnetically charged black holes satisfying the quantisation condition, do not admit a flat limit.

However, we observe that in \cite{Gnecchi:2014cqa} the quantisation condition was not imposed on the specific example of the  $t^3$- magnetic non-extremal solution. In this case, the solution \textit{does} admit a reasonable flat limit. We look into this solution further in Appendix \ref{sec:non-ext.}. Note also that this reference imposes the additional constraint $p_{\Lambda}g^{\Lambda}=\pm1$ when considering BPS magnetic solutions.

\subsection{Extremal, non-BPS solutions}\label{sec:ext.nonBPS}

As an example of the mapping between non-BPS and BPS solutions coupled to vector multiplets, we look at the simple constant scalar electric solution of \cite{Toldo:2012ec}. The metric is of the form
\begin{equation}\label{metric}
    ds^2=U^2(r)dt^2-\frac{dr^2}{U^2(r)}-h^2(r)(d\theta^2+\sin^2\theta d\varphi^2)\,,
\end{equation}
with 
\begin{equation}
U^2(r)=\frac{3\sqrt{3}(\frac{4}{27}\xi_0\xi_1^3 g^2r^2+1-\frac{\mu}{r}+\frac{Q}{r^2})}{2\sqrt{\xi_0\xi_1^3}}\;,\;\;\;\;\;\;\;\;\;\;\;\;\;\;\;h(r)=\frac{\sqrt{2}(\xi_0\xi_1^3)^\frac{1}{4}r}{3^{\frac{3}{4}}}\,,
\end{equation}
and $Q$ is fixed in terms of the FI parameters and electric charges as $Q=\frac{q_1^2}{\xi_1^2}$. The extremal solution is obtained with the following:
\begin{equation}
    \mu_{\text{extr}}=\frac{\sqrt{2}}{3g\sqrt{|V_*|}}(\sqrt{1+4g^2|V_*|Q}+2)(\sqrt{1+4g^2|V_*|Q}-1)^\frac{1}{2}\,,
\end{equation}
and the horizon radius is for this case
\begin{equation}
    r^2_h=\frac{-1+\sqrt{1+4g^2|V_*|Q}}{2g^2|V_*|}a^2\,.
\end{equation}
The first thing to notice is that this horizon radius is finite in the flat limit; expanding around $g=0$ gives
\begin{equation}
    r_h^2=Qa^2+\mathcal{O}(g^2)\text{.}
\end{equation}
Moreover, \cite{Toldo:2012ec} explicitly writes for these solutions the difference between the values of $\mu$ defining the extremality and the BPS bounds. Taking the limit $g\to 0$ in their expression we notice that this difference goes to zero:
\begin{equation}
    \Delta\mu=-g^2\frac{4}{3}V_*\left(\frac{q_1}{\xi_1}\right)^3+\mathcal{O}(g^4)\to 0\text{.}
\end{equation}
This means that, for this class of electric solutions with constant scalars, the extremal black hole becomes also BPS in flat space. This serves as an example in supergravity coupled to vector multiplets of how an extremal AdS black hole which is not BPS can become BPS in the flat limit.

Because the scalar fields are constant, we expect the analysis of the flat limit of the on-shell action to carry through as in the minimal RN case, with the electric charge being replaced by the central charge.


\section{Collecting all results}\label{sec:summary}
In this section we collect our relevant data for clarity. In the table below, we summarise the on-shell actions computed for various backgrounds, and their appropriate flat space limits wherever possible.

\begin{table}[H]
\centering
\scriptsize
\renewcommand{\arraystretch}{1.35}
\setlength{\tabcolsep}{3pt}
\begin{tabular}{p{2.4cm} p{5.3cm} p{4.7cm} p{2.6cm}}
\toprule

\textbf{Setup} &
\textbf{$I_{\text{on-shell, AdS}}$} &
\textbf{$I_{\text{on-shell, flat}}$} &
\textbf{Comments} \\

\midrule

AdS-Schwarzschild
&
\eqref{I_total}
&
\eqref{I_on-shell_SAdS}
&
Small vs.\ large
\\[0.5em]

RN-AdS
&
\eqref{I_Emparan}
&
\eqref{flat_Emparan}
&
Goes to RN
\\[1em]

BPS RN-AdS
&
$\infty$
&
---
&
No smooth flat limit (naked singularity)
\\[0.5em]

Ext RN-AdS
&
---
&
\eqref{eq:IRN_ext_flat}
&
Goes to extremal RN
\\[0.5em]

Near-ext RN-AdS
&
\eqref{discpartitionfunct}
&
\eqref{Flatdiscpartition}
&
Goes to near-BPS / near-ext RN
\\[1em]

Kerr-AdS
&
\eqref{KerrAction}
&
\eqref{eq:IKerr_flat}
&
Goes to Kerr
\\[1.2em]

Kerr-AdS BPS
&
\eqref{KerrAction}$|_{\text{BPS}}$
&
$\infty$
&
No flat limit
\\[0.5em]

Kerr-AdS BPS, $a\,\ell_{AdS}$ fixed
&
\eqref{KerrAction}$|_{\text{BPS}}$
&
\eqref{eq:IKerr_flat}$|_{a=0}$
&
Flat limit exists
\\

\bottomrule
\end{tabular}
\caption{Summary of Euclidean on-shell actions and flat-space limits for various black hole backgrounds.}
\end{table}

We also elucidate the question raised in the introduction regarding the mapping of solutions in the flat space limit:\footnote{In this scheme, we write extremal to implicitly mean extremal and non-BPS, which is automatically true for all solutions but the BPS Kerr-Newman-AdS one. This solution is also extremal. The BPS static solutions are implicitly naked singularities. We also specify to static in the second line due to the case of the Kerr black hole in flat space, which is not BPS at extremality.}
\begin{tcolorbox}[
    left=1mm,
    right=1mm,
    top=1mm,
    bottom=1mm
]
\begin{equation}
\begin{aligned}
&\textbf{Minimal $\mathcal{N}=2$ SUGRA:}
\\[1mm]
\text{Non-extremal}
    &\longrightarrow \text{Non-extremal},
\\
\text{Extremal static}
    &\longrightarrow \text{Extremal BPS static},
\\
\text{BPS static}
    &\longrightarrow \text{Singular},
\\
\text{BPS rotating}
    &\longrightarrow
    \begin{cases}
    \text{Singular under }\ell_{\mathrm{AdS}}\to\infty,\\
    \text{Extremal BPS static under \eqref{appropriate_limit}}.
    \end{cases}
\end{aligned}
\end{equation}
\end{tcolorbox}

Based on the examples given in \ref{sec:ext.nonBPS} and Appendix \ref{sec:non-ext.}, for FI matter coupled gauged supergravity we find that the mapping shown above remains valid provided that the Dirac quantisation condition is not imposed. Since this condition is imposed in the BPS limit, BPS black holes coupled to vector multiplets (or excitations above the magnetic BPS vacuum, as phrased in the conclusion of section \ref{sec:quantisation}) generically do not admit a flat space limit. 

\section{Carroll-like limit of ABJM and black hole entropy} \label{sec:Carrol}

We now comment on the possibility of establishing a concrete duality between the flat space limit in the bulk and the Carroll limit of the CFT. Taking the Carroll limit of quantities in ABJM has been looked into, for example \cite{Lipstein:2025jfj} take the Carroll limit of correlators in ABJM by explicitly taking $c\rightarrow 0$ after a suitable rescaling. At the level of the action, this limit is a bit more subtle. Toy models similar to ABJM have been considered in the context of this limit \cite{Bagchi:2024efs,Miskovic:2023zfz}. However this is only relevant for the bosonic sector of ABJM. The fermionic sector was included in the recent paper \cite{Bagchi:2026emg}. The study of Carroll limits of field theories is still an active area of work. There are also attempts to study Carroll limits on objects like correlators, for example \cite{Lipstein:2025jfj} establish a dictionary between the flat space limit of bulk Witten diagrams and the Carroll limit of field theory correlators.

We would now like to establish the same mapping for semiclassical partition functions. So far we have considered the bulk partition functions in several cases and taken their flat space limits (wherever possible). Based on our findings, we now propose a suitable Carroll-like limit for the partition function of ABJM theory on $S^1\times S^2$. This then gives a prediction for the Carollian field theory dual to the flat Kerr black hole.

\subsection*{The setup: BPS Kerr-AdS black hole and ABJM on $S^1\times S^2$}

In order to establish the duality between the flat limit of an AdS black hole and the Carroll limit of the appropriate dual CFT, we require that the field theory partition function be known exactly, at least to leading order in $1/N$. This is of course because the semiclassical approximation of the gravitational partition function demands that the dual field theory be strongly coupled. We will therefore choose a supersymmetric setup, where CFT partition functions can be computed exactly through supersymmetric localisation. See \cite{Nian:2013qwa} for the relevant localisation computation relevant for the class of field theories we will consider here.

Our results so far tell us what is the appropriate BPS black hole to choose in order to establish the desired result - the BPS Kerr-AdS black hole. This solution has an event horizon and a spherical horizon, and it admits a well-defined flat space limit. However, as we have seen in \ref{flat limit Kerr}, this is not the naive $\ell_{AdS}\to \infty$ limit, but rather the limit \eqref{appropriate_limit} in which we simultaneously send $a\to 0$ with $a\,\ell_{AdS}$ fixed. On the other hand, the entropy of this BPS black hole in AdS has been reproduced by the (Legendre transform of the) free energy of ABJM theory on $S^1\times S^2$ in the large $N$ and in a Cardy-like limit, computed in \cite{Nian:2019pxj}. In the remainder of this section, we will review these results and apply the appropriate limit on both sides of the correspondence. More schematically, these limits will correspond to the those represented by the horizontal arrows in Figure \ref{Scheme}, our goal being to establish the correspondence given by the vertical arrow on the right, in this particular BPS setup.

\subsection{The BPS Kerr-Newman black hole entropy in the flat limit}\label{sec:8.1} 

The black hole solution we will consider is a multi-charge generalisation of the Kerr-Newman-AdS black hole studied in section \ref{sec:rotatingBH}. It arises as a solution of $\mathcal{N}=8$ supergravity when truncating the $SO(8)$ gauge group to its $U(1)^4$ Cartan subgroup. The model thus contains four electric charges, denoted here by $Q_{I, BH}$, $I=1,...,4$, which are set such that $Q_{1, BH}=Q_{3,BH}$ and $Q_{2,BH}=Q_{4,BH}$. See \cite{Chow:2013gba,Cassani:2019mms} for the explicit metric and gauge fields. The Kerr-Newman-AdS black hole originally studied in \cite{Caldarelli:1999xj,Caldarelli:1998hg} is obtained when all the charges are set to be equal \cite{Cassani:2019mms}.

We focus on the BPS limit of this black hole. In \cite{Choi:2018fdc} its entropy function was determined, with the black hole entropy computed as
\begin{equation}
    S_{BH}=\frac{\pi}{g^2G_N}\frac{J_{BH}}{\frac{1}{g}\left(2Q_{1,BH}+2Q_{2,BH}\right)}\,,
\end{equation}
where the angular momentum can be expressed in terms of the charges as 
\begin{equation}
    J_{BH}=\frac{1}{2}\left(\frac{2}{g}Q_{1,BH}+\frac{2}{g}Q_{2,BH}\right)\left(-1+\sqrt{1+16 g^4 G_N^2\frac{2Q_{1,BH}}{g}\frac{2Q_{2,BH}}{g}}\right)\,.
\end{equation}

We now consider taking the $g\to 0$ limit with the charges $Q_{1,BH}$, $Q_{2,BH}$ taking a finite value in the limit. According to the discussion in \ref{flat limit Kerr}, this limit is cannot be achieved through only $g\to 0$, but must also involve sending the rotation parameter to zero, $a\to 0$, with $a/g$ fixed. A simple manipulation shows that the entropy is also well defined in this limit; the angular momentum is
\begin{equation}
    \lim_{g\to 0} J_{BH}= 32\,G_N^2\, g\,Q_{1,BH}Q_{2,BH}(Q_{1,BH}+Q_{2,BH})+\mathcal{O}(g^2)\,,
\end{equation}
and the black hole entropy becomes
\begin{equation}\label{limSBH}
    \lim_{g\to 0}S_{BH}=16\pi G_NQ_{1,BH}Q_{2,BH} \,.
\end{equation}
In particular, the BPS relation remains finite in the limit we consider, and simply takes the form of the BPS relation of ungauged minimal supergravity:
\begin{equation}
    M=\sum_IQ_I\,.
\end{equation}

We would also like to check that \eqref{limSBH} reproduces the area law for the flat space RN black hole if we set all charges equal; this is done in Appendix \ref{App:Q1Q2}. In the limit of equal charges $Q_{1,BH}=Q_{2,BH}=Q_{BH}$, \eqref{limSBH} becomes the entropy of the BPS extremal RN black hole in flat space with 4 equal charges, $S_{BH}^\text{RN, flat}=16\pi G_N Q_{BH}^2$.

\subsection{The appropriate Carroll limit of the ABJM free energy}

Our goal is now to define a dual limit to the flat space limit of the $U(1)^4$ Kerr-Newman black hole discussed in the previous section. The entropy of this black hole has been reproduced in \cite{Nian:2019pxj} from the free energy of ABJM on $S^1\times S^2$ in a Cardy-like limit and in the large $N$ limit. They considered a particular fibration of $S^2$ over $S^1$ of the form
\begin{equation}\label{S1S2}
    ds^2=d\tau^2+L^2\left[d\theta^2+\sin^2\theta(d\varphi-i\Omega d\tau)^2\right]\,,\,\,\,\,\,\,\,\,\,\tau\,\sim\,\tau+L\,,
\end{equation}
and hence $\Omega$ may be interpreted as an angular velocity of an observer rotating along the $\varphi$ cycle; the final result will depend on the combination $\omega\equiv L\Omega$. Their result for the free energy $F=-\log Z$ of ABJM on this background is
\begin{equation}\label{F}
    F\simeq \frac{2\sqrt{2}ik^{\frac{1}{2}}N^{\frac{3}{2}}}{3}\frac{\sqrt{\Delta_1\Delta_2\Delta_3\Delta_4}}{\omega}\,.
\end{equation}
This result is understood as the free energy in the grand-canonical ensemble. Here $\Delta_I$ and $\omega$ are chemical potentials conjugate to the charges, and angular momentum respectively, $k$ is the Chern-Simons level, $N$ is the rank of the gauge group. The free energy is dimensionless overall. Before taking a "Carroll" limit, we first recognise that the theory is Euclidean, and therefore has Euclidean time $\tau$, in which no speed of light can be restored back in the metric. Therefore, the procedure would be to take an appropriate limit analogous to sending $c\rightarrow0$, by keeping quantities fixed, which we clarify concretely in both the canonical and grand-canonical ensemble. In the canonical ensemble, we work with the Legendre transform  of \eqref{F} which is,
\begin{equation}\label{eq:entropy_fct}   
S\left(\Delta_I, \omega\right)=-\frac{2 \sqrt{2} i k^{\frac{1}{2}} N^{\frac{3}{2}}}{3} \frac{\sqrt{\Delta_1 \Delta_2 \Delta_3 \Delta_4}}{\omega}+2 \omega J+\sum_I \Delta_I Q_I+\Lambda\left(\sum_I \Delta_I-2 \omega-2 \pi i\right)\,.
\end{equation}
Here $\Lambda$ acts as a Lagrange multiplier which imposes the constraint $\sum_I\Delta_I-2\omega-2\pi i=0$ upon minimizing $S$ with respect to $\Delta_I$ and $\omega$.

\subsubsection{Carroll-like limit in the canonical ensemble}

We first consider working in the canonical ensemble, with electric charges and angular momentum held fixed. The crucial step in matching the (Legendre transform of the) free energy to the entropy of the BPS black hole in \cite{Nian:2019pxj} was the translation between the parameters of the field theory $Q_I$, $J$ and $k^\frac{1}{2}N^\frac{3}{2}$, and the black hole parameters $Q_{I,BH}$ and $J_{BH}$ and $G_N$, where $Q_I$ and $J$ are conjugate charges to $\Delta_I$ and $\omega$ in an appropriately defined entropy function of the field theory. Let us repeat this dictionary here:
\begin{equation}\label{dictionary}
    \frac{\ell^2_{AdS}}{G_N}=\frac{2\sqrt{2}}{3}k^\frac{1}{2}N^\frac{3}{2}\,,\,\,\,\,\quad Q_{I,BH}=\frac{g}{2}Q_{I,\,\text{ABJM}}\,,\,\,\,\,\quad  J_{BH}=J_{\text{ABJM}}\,.
\end{equation}

This immediately implies the following limit of the charges appearing in the entropy function of the field theory:\footnote{The subscript "ABJM" was included so far to make it clear that we are referring to the charges defined in the field theory, but will be dropped henceforth.}
\begin{equation}\label{abjmlimit}
    J_{\text{ABJM}}\to 0\,,\,\,\,\,\,\,\,\,\,\,\,\,\,\,Q_{I,\,\text{ABJM}}\to\infty\,,\,\,\,\,\,\,\,\,\,\,\,\,\,\,\,\frac{2\sqrt{2}}{3}k^{\frac{1}{2}}N^{\frac{3}{2}}\to \infty\,.
\end{equation}
This occurs in such a way that the Legendre transform of the free energy, after being maximised with respect to the chemical potentials $\Delta_I$ and $\omega$, stays fixed when the limit of interest is taken. This is
\begin{equation}\label{entropyabjmflat}
    S(Q_I,J)=\frac{2\sqrt{2}\pi k^{\frac{1}{2}}N^{\frac{3}{2}}}{3}\frac{J}{Q_1+Q_2}\, ,
\end{equation}
which, by construction, becomes \eqref{limSBH} when the limit is taken, and after translating from field theory charges to bulk charges. Using the constraint,
\begin{equation}
J=\frac{1}{2}\left(Q_1+Q_2\right)\left(-1 \pm \sqrt{1+\frac{18 Q_1 Q_2}{k N^3}}\right),
\end{equation}
and choosing the positive branch, setting the charges equal $Q_1=Q_2=Q$ and expanding for large $N$,
\begin{equation}\label{JQ}
    J=Q\left(\frac{9Q^2}{kN^3}+\mathcal{O}(N^{-6})\right).
\end{equation}
This also agrees with the limits defined in \eqref{abjmlimit}, as $J_{\text{ABJM}}\to0$, when $Q_{I,\,\text{ABJM}}\to\infty\,\,\,\,k^{1/2}N^{3/2}\to\infty$, keeping the combination $Q^2/N^{3/2}$ fixed. Putting it back in \eqref{entropyabjmflat},
\begin{equation}
    S(Q)=\frac{2\sqrt{2}\pi k^{\frac{1}{2}}N^{\frac{3}{2}}}{3}\frac{9Q^3}{kN^3}\frac{1}{2Q}=\frac{3\sqrt{2}\pi Q^2}{k^{1/2}N^{3/2}}.
\end{equation}
Now we can define a cleaner limit that only includes $Q$ and $k^{1/2}N^{3/2}$. Then we take the limit $Q\to\infty,\,\, k^{1/2}N^{3/2}\to\infty$, keeping $\frac{3\sqrt{2}}{k^{1/2}N^{3/2}}Q^2\equiv \tilde{Q}^2$  fixed, which then gives the Carroll-like entropy,
\begin{tcolorbox}
\begin{equation}\label{rescaled_S_canonical}
  S(\tilde{Q})_{\text{Carroll}} = \pi\tilde{Q}^2\,.
\end{equation}
\end{tcolorbox}
\noindent The rescaling of $Q$ we have performed is consistent with $J\to 0$ in the limit under consideration, as seen from \eqref{JQ}.

We now concretely check that this result reproduces the entropy of the flat space limit of the BPS rotating black hole we started with in \ref{sec:8.1}, which is just the BPS and extremal flat space RN black hole with 4 equal charges. For this we again use the dictionary \eqref{dictionary} in order to translate the new charge $\tilde{Q}$ in terms of $Q_{BH}$, resulting in
\begin{equation}
    Q^2_{BH}=\frac{g^2}{4}Q^2=\frac{g^2}{4}\frac{k^\frac{1}{2}N^\frac{3}{2}}{3\sqrt{2}}\tilde{Q}^2=\frac{1}{G_N16}\tilde{Q}^2\,.
\end{equation}
Using this we immediately observe
\begin{equation}
    S(\tilde{Q})_{\text{Carroll}}=\pi\tilde{Q}^2=16\pi G_NQ_{BH}^2=S_{BH}^{\text{RN, flat}}\,,
\end{equation}
matching \eqref{limSBH} for $Q_{1,BH}=Q_{2,BH}$. 

We would like to remark that what we have just presented may be regarded as a microscopic derivation of the entropy of an asymptotically flat, BPS and extremal black hole. By specifying the general construction illustrated in Figure \ref{Scheme} to the supersymmetric case treated here, we were able to identify a limit on the field theory charges and entropy function which reproduces the black hole entropy. We emphasise that this can be done purely from field theory parameters. This is complementary to traditional microstate counting techniques \cite{Strominger:1996sh,Maldacena:1997de} which resort to near-horizon $AdS_3$ region of a given brane system and the associated Cardy formula for the dual CFT$_2$. Here, in our derivation, we did not use the near horizon AdS geometry, but instead the properties of the dual field theory at null infinity.

\section{Discussion}

\subsection{Concluding remarks}

In this work we studied, in the context of $AdS/CFT$ and flat space holography, the flat limit of black hole partition functions in $4d$ minimal and non-minimal gauged supergravity. In particular, we asked whether asymptotically flat black hole solutions can be understood as arising from the flat space limit of asymptotically AdS black holes. 

To study these questions in detail, we started from bulk $AdS_4$ black hole solutions in gauged supergravity and investigated the flat space limits of their (semiclassical) partition functions by computing their on-shell actions and taking $\ell_{AdS}\to\infty$ wherever well-defined. We clarified important aspects of extremal and near-extremal black hole entropy in taking these limits. We  found that some BPS black holes do not admit a regular flat space limit; these solutions are the magnetic AdS black holes, and their flat limits are inhibited by the Dirac quantisation condition. However, we found that the Kerr-Newman-$AdS_4$ BPS solution does have a well defined flat limit when the combination $a\,\ell_{AdS}$ is kept fixed, or equivalently the electric charges are held fixed while angular momentum is turned off. Then, by studying the free energy and entropy in the canonical example of the dual ABJM theory on $S^1\times S^2$ \eqref{F}, we proposed a controlled and well-defined flat limit, which we identify as a Carrollian-like limit. With this limit, illustrated in \eqref{abjmlimit}, we identified the entropy for the corresponding Carrollian theory as in \eqref{rescaled_S_canonical}, which takes the form $S_{\text{Carroll}}=\pi\tilde{Q}^2$, where $\tilde{Q}$ is interpreted as an $R$-charge in the corresponding supersymmetric Carrollian theory \cite{Bagchi:2026emg}. Our result can also be regarded as a microscopic derivation of the entropy of a BPS and extremal black holes in asymptotically flat space from a Carrollian-like perspective. 

Note that the genuine bulk dual of ABJM is $AdS_4\times S^7$ and includes the sphere part, which the free energy \eqref{F} already accounts for through its dependence on the chemical potentials conjugate to the R-symmetry charges. In this paper we have considered only the properties of the four-dimensional asymptotically $AdS_4$ geometry. A crucial consideration is the effect of the flat space limit on the full 11-dimensional geometry, which includes the $S^7$ that decompactifies in the limit. How to formulate Carrollian holography in top-down scenarios when an internal manifold decompactifies is a question under active investigation.\footnote{Recent references provide some insight on this issue: \cite{Bagchi:2026bka} suggests that string scattering in $\mathbb{R}^{1,9}$ can be captured by a double-scaling limit of $\mathcal{N}=4$ SYM on a four dimensional null boundary $\mathbb{R}^{1,3}$; meanwhile, \cite{Eden:2026ulx} shows that certain boundary correlators in $AdS_5\times S^5$ containing contributions from infinite towers of KK modes on the sphere can be recast as correlators whose kinematics is restricted to five dimensions. For another perspective on this issue, see e.g. \cite{Fontanella:2025tbs}.}


There are still many things to be understood about the Carrollian limit of $3d$ ABJM theory \cite{Bagchi:2026emg,Bagchi:2024efs}. 
Our approach could be extended to the grand canonical ensemble, where one holds fixed the chemical potentials. In this case one would have to consider that the parameter $L$, entering in the potential $\omega$ conjugate to the angular momentum in the ABJM theory, is the radius of the two-sphere at the boundary of $AdS_4$, see \eqref{S1S2}. It should therefore be identified with $\ell_{AdS}$ and taken to infinity in the limit we consider. The interplay between the limit we take and the constraint on the chemical potentials, $\sum_I\Delta_I-2\omega=2\pi i\,(\text{mod }4\pi i)$, as well as the Cardy limit $\omega\ll 1$,  are points that should also be investigated further.\\

We conclude with brief comments about two questions that are addressed by our work, namely the consistency of the flat space limit at the level of the counterterm action (before evaluating the on-shell action on a solution); and the possibility of studying the flat space limit from asymptotically de Sitter solutions. 

\subsection{Holographic renormalisation in asymptotically flat spacetimes}

In this work, we have calculated the flat space on-shell actions and/or entropies by taking the flat space limit $\ell_{AdS}\to\infty$ of the on-shell actions defined in asymptotically AdS backgrounds. Specifically, we computed the on-shell actions for asymptotically AdS backgrounds using holographic renormalisation, with the counterterm prescription from \cite{Emparan:1999pm}. Alternatively, we could have also computed the on-shell actions via holographic renormalisation in asymptotically flat spacetimes \cite{Mann:2005yr,Astefanesei:2006zd} of the form,
\begin{equation}
    d s^2=\left(1+\frac{2 \sigma}{\rho^{d-3}}+\mathcal{O}\left(\rho^{-(d-2)}\right)\right) d \rho^2+\rho^2\left(h_{i j}^0+\frac{h_{i j}^1}{\rho^{d-3}}+\mathcal{O}\left(\rho^{-(d-2)}\right)\right) d \eta^i d \eta^j.
\end{equation}
Here, $\rho$ is the radial coordinate with asymptotically Minkowski coordinates $x^a$, related via $\rho^2=\eta_{ab}x^ax^b$. $h_{i j}^0 \text { and } \eta^i$ are both metrics and associated coordinates on the $(d-2,1)$ unit $\mathcal{H}^{d-1}$ hyperboloid. Examples of suitable, finite counterterms were found, one of them depending on the the square-root of the Ricci scalar, 
\begin{equation}
    S_{\sqrt{\mathcal{R}}}=\frac{1}{8 \pi G} \sqrt{\frac{n}{n-1}} \int_{\partial M} \sqrt{-h} \sqrt{\mathcal{R}}\,.
\end{equation}
We ask the reader to turn to \cite{Mann:2005yr} for a fuller treatment.

Although we have not systematically analysed the differences between the results obtained from flat space counterterms and the flat limit of AdS counterterms, we can confirm that both approaches yield the same result for the case of rotating black holes. In fact, \cite{Astefanesei:2006zd} computed the counterterms for specific examples including the asymptotically flat Kerr background, finding that the on-shell action is $I=\beta M/2$, where $M$ is the ADM mass. A quick check tells us that this matches with the flat limit of the on-shell action of the rotating AdS black hole studied in section \ref{sec:rotatingBH}: equation \eqref{KerrAction} in the flat space limit can be written as $I_{\text{flat, Kerr}}=\beta m/2G$ using the relation,
\begin{equation}
   m=\frac{1}{2}\left(r_h+\frac{a^2}{r_h}\right),
\end{equation}
which is obtained by solving for $r_h$ in \eqref{paramskerr}.

We also note that \cite{Mann:1999pc} has written a counterterm that interpolates between that of AdS for small $\ell_{AdS}$ and flat space for $\ell_{AdS}\to\infty$, in four dimensions.

\subsection{Flat space limit from de Sitter}

In this work we have approached flat space holography by taking a limit from $AdS_4$. A reasonable question is to study the flat space limit starting from $dS_4$, but which turns out to be more subtle. We can compute the on-shell action of the Schwarzschild-de Sitter black hole and the extremal Nariai solution and take the naive flat space limit. In fact, the on-shell action has been computed in \cite{Morvan:2022aon}. Their result is
\begin{equation}\label{morvan}
    I_{SdS}=-\frac{A_b+A_c}{4G}\,,
\end{equation}
where $A_b$ and $A_c$ are the areas of the black hole and cosmological horizons respectively.
This is related to the result obtained by Bousso and Hawking \cite{Bousso:1996au},
\begin{equation}\label{boussohawking}
    I_{SdS}=-\frac{\mathcal{V} \Lambda}{8 \pi G}-\frac{A_b \epsilon_b}{8 \pi G}-\frac{A_c \epsilon_c}{8 \pi G}\,,
\end{equation}
where $\mathcal{V}$ is the four-volume of the geometry and  $\epsilon_{b,c}$ are the conical deficit angles. The difference arises from using the Smarr relation to go from \eqref{boussohawking} to \eqref{morvan}.  

For generic Schwarzschild--de Sitter geometries the two horizons have different temperatures, $(T_b\neq T_c)$, and hence no single choice of Euclidean periodicity can simultaneously remove both conical singularities. Such configurations should therefore be regarded as constrained instantons \cite{Morvan:2022aon} rather than ordinary thermal saddles. The two limiting configurations are special: empty de Sitter and the Nariai geometry are regular Euclidean saddles, with actions
\begin{align}
I_{dS}=-\frac{3\pi}{\Lambda G},
\qquad
I_{\mathrm{Nariai}}=-\frac{2\pi}{\Lambda G}.
\end{align}
As the black-hole mass is increased from $M=0$ to the Nariai value $(M=M_N)$, the constrained $SdS$ action varies continuously between these two values. Equivalently, the total horizon entropy decreases,
\begin{equation}
S_{dS}>S_{SdS}>S_{\mathrm{Nariai}},
\end{equation}
so that increasingly large $SdS$ black holes are progressively suppressed relative to empty de Sitter.

Let us now comment on the flat space limit. For a "large" black hole in $dS_4$, we consider the Nariai solution which is the largest black hole you can have where $r_b=r_c\equiv r_N$. Immediately we see that it gives a diverging contribution to the partition function, just like large black holes in $AdS_4$. A "small" black hole in de Sitter would correspond to a $SdS_4$ black hole.\footnote{We mean small and large as analogous classifications to $AdS_4$ black holes.} Since there are two area contributions to the action, the flat space limit pushes the cosmological horizon $r_c$ to infinity, giving a divergent piece. In essence,
\begin{align}
    I_{SAdS_4}&\overset{\ell_{AdS}\to\infty}\longrightarrow I_{\text{flat}}\,,
    \\
    I_{SdS_4}&\overset{r_c\to\infty}\longrightarrow I_{\text{flat}}+\text{divergent}.
\end{align}
Note that the divergent piece does not come from the $SdS_4$ black hole, but  from the $dS_4$ contribution. Therefore we can instead consider the "renormalised" quantity,
\begin{equation}\label{gamma}
    \Delta I=I_{SdS_4}-I_{dS_4}\,,
\end{equation}
where we subtract the empty $dS_4$ background. With this prescription, the flat space limit from $AdS_4$ and $dS_4$ yield the same $I_{\text{flat}}$. The quantity \eqref{gamma} is also related to pair creation rate \cite{Bousso:1996au} where,
\begin{equation}
    \Gamma_{\text{pair}}\sim e^{-({I_{SdS}-I_{dS}})}\sim\frac{Z_{SdS}}{Z_{dS}}
\end{equation}
computes the probability/rate at which a black hole nucleates from $dS_4$.

Finally, unlike AdS black holes, generic SdS black holes do not constitute a family of configurations in global thermal equilibrium: the regular endpoints are empty de Sitter and Nariai, while intermediate SdS geometries have unequal horizon temperatures and enter the Euclidean path integral as constrained instantons.

In conclusion, the semiclassical partition function of Schwarzschild black holes can be seen as arising from the flat space limit of either SAdS or SdS black holes, after the divergence of the cosmological horizon is removed according to \eqref{gamma}. This constitutes the first step in showing the complementarity of our results in AdS with the flat limit from dS.

The question regarding flat space limits from AdS and dS was also considered in \cite{Hidalgo:2026ggx} in two boundary dimensions, where flat limits of $AdS_3$ and $dS_3$ were studied. It would be interesting to connect the two approaches.

\section*{Acknowledgements}

We thank Romain Ruzziconi, Chiara Toldo and Roberto Emparan for useful comments. P.V.M.\ is supported by the Netherlands Organisation for Scientific Research (NWO) under the VICI grant VI.C.202.104.
\appendix 
\section{Non-extremal black holes with running scalars and flat limit}\label{sec:non-ext.}

To illustrate our discussion in section \ref{sec:quantisation} and how the flat space limit operates with running scalars, we now consider taking the flat space limit of black hole solutions for which the quantisation condition is not imposed, verifying that the flat limit indeed exists.
The on-shell action of black holes in $4d$ gauged supergravity have been computed explicitly in, for example, \cite{Halmagyi:2017hmw} and \cite{Gnecchi:2014cqa}. We collect the results from \cite{Gnecchi:2014cqa}, focusing on the non-extremal electric and magnetic solutions of $\mathcal{N}=2$ supergravity with FI gaugings and the prepotential $F=-2i\sqrt{X^0(X^1)^3}$. The solutions therefore contain a single scalar $z=\frac{X^1}{X^0}$. The metric again takes the form \eqref{metric}, with $U(r)^2=e^Kf(r)$ and $h(r)^2=e^{-K}r^2$. The function $f(r)$ takes the generic form
\begin{equation}
    f(r)=\kappa + \frac{c_1}{r}+\frac{c_2}{r^2}+\tilde{g}^2r^2e^{-2K(r)}\,,
\end{equation}
where $\tilde{g}=\frac{\sqrt{2}\xi_0^{1/4}\xi_1^{3/4}}{3^{3/4}\beta}g$, with $\beta$ and overall factor in $e^{-K(r)}$ appearing as
\begin{equation}
    e^{-K(r)}=\beta^2\sqrt{H_0(H_1)^3}\,,\,\,\,\,\,\,\,\,\,\,H_\Lambda=a_\Lambda+\frac{b_\Lambda}{r}\,,\,\,\,\,\,\Lambda=0,1\,.
\end{equation}

We now quote the results for the renormalized on-shell actions of magnetic and electric black holes in this model, with a scalar potential chosen as $V(\varphi)=-\frac{3}{\ell_{AdS}^2}\cosh\left(\sqrt{\frac{2}{3}}(\phi(r)-\sqrt{3/8}\log(3\xi_0/\xi_1))\right)$, with $z=e^{x\phi(r)+y}$ such that $\phi(r)$ is a canonically normalized field.

\begin{equation}
    I^{Mag}_{on-shell}=-\frac{3c_1}{4}-\frac{r_h}{2}+\frac{1}{\ell^2}\left(2Q_1-r_h\right)\left(Q_1+r_h\right)^2
\end{equation}
where 
\begin{equation}
    c_1=\frac{b_0}{a_0}+\frac{b_1}{a_1}+\frac{2}{\beta^2}\frac{a_0a_1}{b_0a_1-b_1a_0}\left(\frac{(p^1)^2}{a_1^2}-\frac{(p^0)^2}{a_0^2}\right)
\end{equation}
and $a_\Lambda=-\sqrt{2}\mathcal{G}^\Lambda\ell_{AdS}$, which can be written as $a_0=\frac{\sqrt{2}\xi_1^{3/2}g\ell_{AdS}}{3\sqrt{3\xi_0}}=\frac{\xi_1}{3\xi_0}a_1$.

For the electric solution we have

\begin{equation}
       I^{Elec}_{on-shell}=\frac{1}{2}\left(\frac{c_1}{2}+r_h\right)
\end{equation}
with
\begin{equation}
    c_1=\frac{b_0}{a_0}+\frac{b_1}{a_1}+\frac{2}{3\sqrt{3}\beta^2}\frac{a_0a_1}{b_0a_1-b_1a_0}\left(\frac{(q_1)^2}{a_1^2}-\frac{(q_0)^2}{a_0^2}\right)
\end{equation}
and this time $a_\Lambda=\frac{\sqrt{2}}{3^{3/4}}\ell_{AdS}g_\Lambda$. We now take the flat space limit of the magnetic and electric on-shell action. It is easy to see for the magnetic case that this just becomes,
\begin{equation}
\begin{aligned}
    I^{Mag}_{on-Shell}\overset{\ell\to\infty}=& -\frac{3c^{\text{flat}}_1}{4}-\frac{r^{\text{flat}}_+}{2}
\\
    &=-\frac{3}{4a_1}\left[\frac{3\xi_0b_0}{\xi_1}+b_1+\frac{2}{3\sqrt{3}\beta^2}\frac{\xi_1/3\xi_0}{b_0-b_1\xi_1/3\xi_0}\left((p^1)^2-(3\xi_0/\xi_1)^2(p^0)^2\right)\right]-\frac{r^{\text{flat}}_+}{2}
\end{aligned}
\end{equation}
For the electric,
\begin{equation}
    \begin{aligned}
    I^{Elec}_{on-Shell} \overset{\ell\to\infty}=& \frac{3c^{\text{flat}}_1}{4}+\frac{r_h^{\text{flat}}}{2}
\\
    &=\frac{1}{4a_1}\left[\frac{\xi_1b_0}{\xi_0}+b_1+\frac{2}{\beta^2}\frac{\xi_0/\xi_1}{b_0-b_1\xi_0/\xi_1}\left((q_1)^2-(\xi_1/\xi_0)^2(q_0)^2\right)\right]+\frac{r^{\text{flat}}_+}{2}
\end{aligned}
\end{equation}
A further check would be to compare these limits with the on-shell action of the asymptotically flat STU black hole of $\mathcal{N}=2$ ungauged supergravity.

\section{Area law for the flat limit of the $U(1)^4$ Kerr-Newman black hole}\label{App:Q1Q2}

We make use of expressions in \cite{Chow:2013gba,Cassani:2019mms} for the electric charges in terms of the parameters appearing explicitly in the metric, namely $\delta_I$, $m$, $a$ and $g$, where $\delta_I$ determine the electric charges,\footnote{Recall that we are setting $Q_{1,BH}=Q_{3,BH}$ and $Q_{2,BH}=Q_{4,BH}$.} $m$ is a mass parameter and $a$ a rotation parameter. In particular, we will use constraints among these parameters imposed by the BPS limit. 

The product $Q_{1,BH}Q_{2,BH}$ appearing in the flat limit of the black hole entropy in \eqref{limSBH} is
\begin{equation}   \label{Q1Q2} Q_{1,BH}Q_{2,BH}=\frac{m^2s_1c_1s_2c_2}{4(1-a^2g^2)^2}\,,
\end{equation}
with $c_I\equiv\cosh(\delta_I)$, $s_I\equiv \sinh(\delta_I)$. Specifying to the BPS limit, we have that the solution is supersymmetric if
\begin{equation}\label{ag}
    ag=\frac{2}{e^{2(\delta_1+\delta_2)}-1}\,,
\end{equation}
while a regular horizon exists for the value of $m$
\begin{equation}
    m_*^2=\frac{1}{g^2}\frac{\cosh^2(\delta_1+\delta_2)}{4e^{\delta_1+\delta_2}\sinh^3(\delta_1+\delta_2)s_1c_1s_2c_2}\,.
\end{equation}

As in section \ref{flat limit Kerr}, we now want to impose the flat space limit $\ell_{AdS}\to\infty$, with $a\to 0$ and $a\ell_{AdS}=ag^{-1}$ fixed. In order to translate this into a condition on the parameters $\delta_I$, we may look at \eqref{ag}, 
\begin{equation}    \label{agto0}
    ag\to 0\,\Rightarrow \,e^{2(\delta_1+\delta_2)}\to\infty\,.
\end{equation}
In particular, we may replace $e^{2(\delta_1+\delta_2)}\sim\frac{2}{ag}$ in the limit. Treating all electric charges in the same way, this also implies $e^{2\delta_I}\sim\sqrt{\frac{2}{ag}}$. Plugging this into \eqref{Q1Q2}, we obtain in the limit
\begin{equation}
    Q_{1,BH}Q_{2,BH}\to \frac{1}{8g^2}e^{-2(\delta_1+\delta_2)}\sim\frac{1}{16}a\ell_{AdS}\,,
\end{equation}
where it is understood that $a\ell_{AdS}$ equals a finite constant in this limit. Plugging this into \eqref{limSBH} one obtains $S_{BH}\to \pi a\ell_{AdS}$.\footnote{Since reference \cite{Cassani:2019mms} sets $G_N=1$ in the expressions we are quoting, we shall do the same in this Appendix for ease of comparison.} In order to check that this indeed corresponds to the area law, note that the area of the horizon is given by $A=4\pi (r_1r_2+a^2)$, with $r_I=r_*+2m_*s_I^2$. In the BPS limit, we have $r_*\equiv \frac{2ms_1s_2}{\cosh(\delta_1+\delta_2)}$, which can be seen to vanish under \eqref{agto0}. This implies
\begin{equation}
    \frac{A}{4}\to 4\pi m_*^2s_1^2s_2^2\sim 4\pi~(2a^2)\frac{e^{2(\delta_1+\delta_2)}}{16}\sim \pi a\ell_{AdS}=S_{BH}\,.
\end{equation}
This concludes our check that the flat space limit of the entropy of the $U(1)^4$ Kerr-Newman-AdS black hole is well-defined and, if all charges are set equal, reproduces the area law for the single-charge RN black hole of minimal supergravity.

\bibliography{references}

\end{document}